\documentclass[useAMS, usenatbib, fleqn]{mn2e}

\usepackage{times}
\usepackage{url}
\usepackage{amsmath}
\usepackage{graphicx}
\usepackage{subfig}
\usepackage{float}
\usepackage{caption}
\usepackage{amsfonts}
\usepackage{amssymb}
\usepackage{multirow}
\usepackage{color}
\usepackage[breaklinks, colorlinks, citecolor=blue, linkcolor=black, urlcolor=black]{hyperref}

\newcommand{\photoz}{photo-$z$}
\newcommand{\ie}{\textit{i.e.,~}}
\newcommand{\eg}{\textit{e.g.,~}}
\newcommand{\equref}[1]{Eq.~(\ref{#1})}
\newcommand{\figref}[1]{Fig.~\ref{#1}}
\newcommand{\ang}{{\bmath{n}}}
\newcommand{\pqso}{{\rm P}_{\rm QSO}}

\newcommand{\ellmax}{{\ell_{\rm max}}}
\newcommand{\mCl}{{\mathcal{C}_\ell}}
\newcommand{\Pl}{{\mat{P}^\ell}}

\newcommand{\nsys}{{\rm N_{sys}}}
\newcommand{\nside}{{\rm N_{side}}}
\newcommand{\tr}{{\textrm{Tr}\ }}

\newcommand{\equ}[1]{\begin{equation}#1\end{equation}}
\newcommand{\eqn}[1]{\begin{eqnarray}#1\end{eqnarray}}
\newcommand{\mat}[1]{\mathbfss{#1}}

\newcommand{\bra}{\langle}

\newcommand{\ket}{\rangle}

\renewcommand{\vec}[1]{\bmath{#1}}

\newcommand{\fnl}{f_{\rm NL}}

\usepackage{color}

\pagerange{\pageref{firstpage}--\pageref{lastpage}} \pubyear{2013}
\def\LaTeX{L\kern-.36em\raise.3ex\hbox{a}\kern-.15em
    T\kern-.1667em\lower.7ex\hbox{E}\kern-.125emX}

\title[Optimal power spectra of XDQSOz quasars]
{Exploiting the full potential of photometric quasar surveys: \\ Optimal power spectra through blind mitigation of systematics}

\author[Leistedt and Peiris]
  {Boris~Leistedt and Hiranya~V.~Peiris\\
  Department of Physics and Astronomy, University College London, London WC1E 6BT, U.K \\ 
   Emails: boris.leistedt.11@ucl.ac.uk, h.peiris@ucl.ac.uk}

\begin{document}

\maketitle 

\begin{abstract}
We present optimal measurements of the angular power spectrum of the XDQSOz catalogue of photometric quasars from the Sloan Digital Sky Survey. These measurements rely on a quadratic maximum likelihood estimator that simultaneously measures the auto- and cross-power spectra of four redshift samples, and provides minimum-variance, unbiased estimates even at the largest angular scales. Since photometric quasars are known to be strongly affected by systematics such as spatially-varying depth and stellar contamination, we introduce a new framework of \textit{extended mode projection} to robustly mitigate the impact of systematics on the power spectrum measurements. This technique involves constructing template maps of potential systematics, decorrelating them on the sky, and projecting out modes which are significantly correlated with the data. Our method is able to simultaneously process several thousands of nonlinearly-correlated systematics, and mode projection is performed in a blind fashion. Using our final power spectrum measurements, we find a good agreement with theoretical predictions, and no evidence for further contamination by systematics. Extended mode projection not only obviates the need for aggressive sky and quality cuts, but also provides control over the level of systematics in the measurements, enabling the search for small signals of new physics while avoiding confirmation bias.
\end{abstract}

\section{Introduction}

Quasars are bright, highly biased tracers of the large scale structure (LSS) of the universe, and are useful for testing cosmological models in large volumes and over extended redshift ranges.  In particular, their bias can inform us about the abundance and mass of dark matter halos in which they form, opening new windows on galaxy formation and the astrophysics of active galactic nuclei (\eg \citealt{Fan:2005es}). In addition, they can be used to constrain primordial non-Gaussianity (PNG) which is predicted to enhance the bias of LSS tracers on large scales \citep{Dalal2008png, matarrese2008, 2009astro2010S158K, loverde2008}. However, these applications require accurate auto- and cross-correlation power spectrum measurements, which are complicated by the presence of significant systematics in the data, and the difficulty of creating large, deep quasar catalogues. Indeed, although quasar candidates are easily confirmed with spectroscopy, quasars are point sources, and most point sources in the sky are stars. Since acquiring high resolution spectra is a time- and resource-consuming process, the creation of large quasar catalogues is cumbersome and can only be realised by preselecting targets and scheduling them for spectroscopic follow-up. In optical frequencies, catalogues of photometric quasars were constructed from the Sloan Digital Sky Survey (SDSS, \citealt{Gunn2006}), and promising candidates were followed up using the SDSS spectrograph, yielding large catalogues of confirmed quasars such as the Baryon acoustic OScillations Survey (BOSS, \citealt{dawson2013boss}).  

While spectroscopic catalogues don't suffer from stellar contamination, photometric samples are larger and extend to fainter magnitudes, and can therefore yield more precise measurements of the clustering and the bias of quasars. However, photometric data are significantly contaminated by multiple sources of systematics, either intrinsic (\eg dust extinction), observational (\eg seeing, airmass), or instrumental (\eg instrument calibration), and star-quasar separation using only colour information is also nontrivial. These systematics affect the properties of the raw images in complex ways, propagate into the final quasar catalogues, and create spurious spatial correlations (spatially-varying star-quasar separation efficiency will induce a spurious clustering signal, see \eg \citealt{Huterer2012calibrationerrors}). Finally, some of these correlations may also be imprinted in spectroscopic catalogues, since the latter rely on targets selected from photometric quasars. Therefore, not just precise --- but also accurate --- cosmological inferences from quasar surveys require careful mitigation of systematics.

The first studies of the clustering of quasars in optical frequencies used both spectroscopic (\eg \citealt{Outram20032dfqso, Shen2007specqso, Ross2009specqsodr5}) and photometric  (\eg \citealt{Myers2006first}) catalogues from early SDSS data, and were used to constrain numerous cosmological and astrophysical quantities of interest, such as the quasar bias, PNG, and the quasar luminosity function \citep{Richards2006qlf, Serber2006, Myers2007one, Strandbrunner2008, SlosarHirata2008}. They exhibited power excesses on large and small scales \citep{Myers2007two}, which were even more significant in the DR6 photometric quasar catalogue \citep{Richards2008rqcat}. Although cuts to the data were not sufficient to remove this excess power, thus pointing to PNG \citep{Xia2009highzisw, Xia2011sdssqsocell, Giannantonio01112012, Karagiannis2013}, recent work has demonstrated that the excess power was due to systematics \citep{PullenHirata2012} and could be eliminated using mode projection  \citep{Leistedt2013excessdr6}. Indeed, it is known that spatially-varying stellar contamination can combine with other systematics and generate such excess clustering power, mimicking significant levels of PNG \citep{Giannantonio2013png}. Alternatively, other studies have focused on using the cross-correlations of the photometric quasars with other data  \citep{Giannantonio2013crosscmblss}, thus extracting some of the information they contain while avoiding the main systematics. The resulting PNG constraints were competitive with those obtained using normal galaxies (\eg \citealt{ross2012png}).  The eighth data release of SDSS yielded a new catalogue of photometric quasars, XDQSO  \citep{Bovy2010xdqso}, relying on the extreme deconvolution technique \citep{Bovy2011xd}; the latest spectroscopic data were used to provide an improved probabilistic selection of quasars, with greater control over completeness issues.  XDQSO extends to fainter magnitudes, and was primarily used to select high redshift candidates scheduled for spectroscopic follow-up in the context of BOSS \citep{Ross2012dr9qsotarget}. Its extension, coined XDQSOz \citep{bovy2012xdqsoz}, provides estimates and probability density distributions for the photometric redshifts of the catalogued objects. XDQSOz was cross-correlated with the cosmic microwave background (CMB) lensing map from the Atacama Cosmology Telescope \citep{ACT2013} to constrain the quasar bias \citep{Sherwin2012qsolensing}. Constraints on PNG were also derived from clustering measurements \citep{agarwalho2013xdqsoz, hoagarwal2013xdqsoz}, but required significant cuts and corrections to exploit measured power spectra, and corrected for systematics using methodologies introduced in \cite{ross2011weights, ross2012systematics, ho2012cosmoweights, agarwalho2013sys}.

Quasar clustering studies therefore remain suboptimal and limited due to cuts and corrections needed to address the high levels of spurious correlations created by systematics. Nevertheless, most of the potential systematics are actually known and can be mapped onto the sky. It is their complex impact on the data which is not known, and prevents the modelling of spurious correlations when estimating power spectra. Mode projection, however, can mitigate the impact of systematics in a robust manner, while minimising the need to throw out hard-won data through masking and cuts \citep{Tegmark:1997yq, THS1998future, SlosarSeljak2004modeproj, PullenHirata2012, Leistedt2013excessdr6}. The purpose of this work is to extend the standard mode projection approach by designing a generic methodology to mitigate the impact of large numbers of known systematics in a flexible and robust manner. We apply this technique to the XDQSOz catalogue in order to precisely control the level of the contamination and accurately measure the power spectrum on the largest scales, which is essential for constraining PNG. Next generation photometric surveys, such as the Dark Energy Survey\footnote{\url{www.darkenergysurvey.org}} (DES), will reach unprecedented precision and will require careful treatment of systematics. This {\it extended mode projection} approach will enable the full potential of such surveys to be exploited in the search for new physics.

This article is organised as follows. In Sec. 2 we recall the definitions and properties of quadratic power spectrum estimators, and introduce the extended mode projection technique to mitigate systematics when estimating power spectra. In Sec. 3 we turn to the XDQSOz catalogue of photometric quasars. We present our data samples, theory predictions and power spectrum measurements, and discuss the ability of the extended mode projection, in combination with blind analysis techniques, to mitigate systematics. The discussion and conclusions are presented in Sec. 4.

\section{Theory and methods}

\subsection{Power spectrum estimation}

The statistics of a Gaussian random field on the sphere are entirely characterised by its power spectrum $C_\ell$.  Any realisation of this field, denoted by $x$, has an observed power spectrum $\mCl$, defined as the variance of the spherical harmonic coefficients $a_{\ell m}$, 
\equ{
	\mCl  \ = \   \sum_{m = -\ell}^{\ell} \frac{ |a_{\ell m}|^2 }{2\ell + 1} \label{harmps},
}	
which, {\it on the full sky only}, is an unbiased estimator of $C_\ell$ (\ie over many realisations, $\bra \mCl \ket = C_\ell$) with cosmic variance
\equ{
	{ {\rm Var}(\mCl)  }{}= \frac{2C^2_\ell}{2\ell+1} \label{cosmicvar}.
}	

Cosmological models usually predict theory $C_\ell$s which can then be confronted with observed $\mCl$s to constrain model parameters.  However, when measuring $\mCl$ with real data, several issues arise. Most datasets are only defined on a portion of the sky, and $\mCl$ must then be estimated from a cut-sky -- or masked -- data vector $\vec{x}$. This issue has been extensively studied in the context of the CMB (\eg \citealt{Teg97, KBJ98, Efsta2003, Efsta2004, Efsta2006, PP10, GruetjenAndShellard2012}), and also more  specifically for LSS data (\eg \citealt{Tegmark2002earlysdss, ho2012cosmoweights, Leistedt2013excessdr6}). The two main quadratic angular power spectrum estimators are the pseudospectrum (PCL) and quadratic maximum likelihood (QML) estimators. In both cases, band-power estimates $\hat{C}_b$ (\ie estimates of $\mCl$ in multipole bins $\Delta\ell_b$) are quadratic in the data,
\equ{
	\hat{C}_b = \vec{x}^t \mat{E}^b \vec{x},
}
bl{where $\mat{E}^b$ characterises the estimator under consideration.} The covariance of the estimates, denoted by $\mat{V}_{bb'} = {\rm Cov}(\hat{C}_b, \hat{C}_{b'})$, includes a cut-sky induced variance which is larger than the cosmic variance and specific to the estimator at hand, and a contribution from the noise and systematics in the data. The equations and implementation details we use are detailed in \cite{Leistedt2013excessdr6}, and in this paper we only provide a summary of the main characteristics of these estimators. 

The PCL estimator deconvolves the mask-induced mode-coupling to produce unbiased estimates of $\mCl$. The estimator is straightforward to implement, not very computationally intensive, but only optimal (\ie it provides minimum-variance estimates) when $\mCl$ is close to flat, and if the mask has a simple geometry. The QML estimator is unbiased in $\mCl$ and is a minimum variance estimator, but is computationally intensive and requires priors on the power spectrum, noise and potential systematic uncertainties in the data. These priors go in a pixel-pixel covariance ${\mat{C}}$  modelled as the superposition of a signal part ${\mat{S}}$ calculated using a theory prior $\{ C_\ell \}$ and noise $\mat{N}$, \ie
\equ{
	\mat{C} = \bra \vec{x}\vec{x}^t \ket = \mat{S} + \mat{N} = \sum_\ell  C_\ell \Pl+ \mat{N}, \label{covarmatrix}
} 
where $(\Pl)_{ij} = ({2\ell+1})/{4\pi}\ P_\ell(\ang_i\cdot\ang_j)$ is a useful matrix notation \citep{Teg97} in which $\ang_i$ denotes the angular position of the $i$th pixel.  In fact, it can be shown that PCL is equivalent to QML when assuming uncorrelated pixels, \ie using a flat power spectrum as prior. Further complications arise because power spectra of real data can only be accurately estimated in a multipole range $0< \ell<\ellmax$, where $\ellmax$ mainly depends on the noise level and the resolution at which the pixelised maps are analysed. These parameters must be consistently adjusted when constructing priors and computing the power spectrum estimates. Rules that guarantee the correct implementation of the PCL and QML estimators for LSS studies are detailed in \cite{Leistedt2013excessdr6}.

\subsection{Extended mode projection} \label{sec:extmodeproj}

As mentioned in the introduction, real data are often contaminated by systematic effects which create spurious correlations in the measured power spectra. A first order solution is to construct masks to ignore the regions of the sky which are not reliable, for example by rejecting pixels based on quality cuts, or where potential systematics are significant (\eg dust, seeing). However, this approach does not remove the spurious correlations due to spatially-varying depth or stellar contamination {\it in the unmasked region}. Hence, masking is insufficient when these systematics are not negligible compared to noise and the cosmic variance. In addition, several {\it ad hoc} choices are required to define what one means by unreliable regions, decide which systematics to include, and what cuts to apply in order to construct the sky mask. As a consequence, there is a risk of being subject to confirmation bias, \ie one may perform the analysis using an initial mask, then notice it is insufficient to remove the impact of systematics, and therefore perform another analysis with a more stringent mask, until the results are in accordance with expectation.  

To illustrate more robust solutions to mitigate systematics, let us start with a toy model where the observed data $\vec{y}$ is the superposition of the systematics-free data vector $\vec{x}$ (of covariance $\mat{C}$) and a set of $\nsys$ systematics maps $\vec{t}_i$, \ie
\equ{
	\vec{y} = \vec{x} + \sum_{i=1}^{\nsys} \alpha_i \vec{t}_i \label{linearcontamination}.
}

With a method to estimate and fix the $\alpha_i$ parameters at hand,  one could correct the maps or the power spectra to remove the impact of the systematics \citep{ross2011weights, ross2012systematics, ho2012cosmoweights}. However, if the systematics are not well-described by a linear model, the correction may be not be robust, and create biases in the measured power spectra. 

Alternatively, one could sample the coefficients $\alpha_i$, estimate the power spectra of the corrected data vector $\vec{y}-\sum \alpha_i \vec{t}_i$, and finally perform a Bayesian Monte Carlo marginalisation over the $\alpha_i$s. This approach can be seen as an analytic marginalisation of systematics to linear order, and is more robust than attempting to correct the data or the power spectra. Interestingly, this approach can be implemented in the QML estimator to directly obtain power spectrum estimates for the systematics-free map $\vec{x}$. This {\it mode projection} (\eg \citealt{Tegmark:1997yq, SlosarSeljak2004modeproj, PullenHirata2012, Leistedt2013excessdr6}) is implemented by using a prior covariance matrix 
\equ{
	\mat{D} = \mat{C} + \lim_{\epsilon \rightarrow \infty} \sum_i \epsilon \vec{t}_i\vec{t}_i^t, \label{modeprojection1} 
}
and setting $\epsilon$ to large values, \ie by giving a large variance to the templates $\vec{t}_i$. Hence, the spatial modes corresponding to these templates\footnote{In this paper, we will use the terms templates and spatial modes interchangeably.} are considered as noise and ignored when estimating the power spectra.

To gain intuition into the mode projection process and find its limitations, let us first notice that the results of the QML estimator are invariant under rotation, \ie transforming the data and the prior covariance matrix $\mat{D}$ using an arbitrary rotation matrix does not affect the band-power estimates (the rotation is a linear transformation of the data pixels and conserves the information content). Therefore, it is straightforward to show that using mode projection with only one template is equivalent to masking one pixel of the data vector $\vec{x}$ in an arbitrary rotated frame (defined by the template). With multiple templates, this masking interpretation is only valid if the templates are orthogonal on the patch of sky of interest, \ie if $\vec{t}_i^t\vec{t}_j = \delta_{ij}$. In this case only, projecting out $\nsys$ templates is equivalent to masking $\nsys$ pixels in some rotated frame. Hence, the increase in variance in the band-power estimates scales as the number of templates. However, real systematics  are often correlated, and projecting out $\nsys$ systematics may correspond to masking less than $\nsys$ pixels, leaving no control over the effective increase in variance in $\mat{V}_{bb'}$. Moreover, without physically-motivated models for the contamination, one may want to include large numbers of templates and consider more generic data-driven models. Yet, in this case many modes may in fact not be relevant in treating systematics, and only contribute to increasing the variance of the band-power estimates. The standard mode projection framework is not appropriate in these situations, because it leaves no control over the increase in variance or the accepted level of contamination when projecting out numerous correlated modes.

To address these issues, we extend the mode projection framework and add two operations to it, prior to power spectrum estimation. Firstly, the systematics templates are decorrelated on the patch of sky of interest. This eliminates the redundant information and leads to a minimal representation of the input templates. Secondly, the resulting orthogonal modes are cross-correlated with the data, yielding null tests that can be used to select and only project out the modes which significantly contaminate the data. This selection can be done according to any criteria of interest, such as cuts based on reduced $\chi^2$ for null tests, which is the metric we use in this work. Together, these two extensions provide control over the effective number of modes and the increase in variance due to mode projection. We now detail our implementation of \textit{extended mode projection} in the context of galaxy and quasar survey analysis. 

For the first step, we stack the systematics templates to construct a matrix $\mat{T}$ where the rows contain the modes to be projected out. As detailed in the next sections about SDSS quasars, we will in fact use modes $\vec{t}_i$ which are either base systematics templates, or products of templates, so that \equref{linearcontamination} will be extended to non-linear contamination models. The templates  $\vec{t}_i$ are then decorrelated using standard singular value decomposition (SVD) techniques to find a transformation matrix $\mat{W}$ which, once applied to $\mat{T}$, yields uncorrelated templates contained in the rows of a matrix $\mat{U}$. This procedure is encapsulated in the set of equations\footnote{Note that one may modify these equations to incorporate a covariance matrix and use the Sherman-Morrison formula to analytically perform the mode projection. In this work, we have employed \equref{modeprojection1} instead, to avoid the need to reweight all templates using a large covariance matrix, which is computationally very intensive.}
\eqn{	
	\mat{T}^t \mat{T} &=&  \mat{W}{\bf \Lambda}\mat{W}^t \\
	\mat{U} &=& \mat{T} \mat{W} \\
	\mat{U}^t \mat{U} &=& \mat{W}^t \mat{T}^t \mat{T}\mat{W} = {\bf \Lambda} \, ,
}
where ${\bf \Lambda}$ is a diagonal matrix containing the eigenvalues $\lambda_i$ of $\mat{T}$. The advantage of this decorrelation step is twofold. Firstly, the output templates, denoted by $\vec{u}_i$, are orthogonal ($\vec{u}_i^t \vec{u}_j = \lambda_i\delta_{ij}$), and yield a minimal description of the input systematics. Secondly, the SVD provides the number and the significance of the orthogonal modes in this minimal description. In particular, it may find modes which are consistent with numerical noise\footnote{A mode is considered as numerical noise if the ratio between its eigenvalue and the largest eigenvalue of the system is smaller than numerical precision.}, and can safely be ignored since they are redundant or not significant in the input system.

The second step aims to select which of these orthogonal modes $\vec{u}_i$ should be projected out. In this work, we resort to $\mCl$ estimators to measure cross-correlation power spectra between the data and the systematics templates. In principle, any measure of correlation could be used, but resorting to power spectrum estimators captures scale information, which is desirable since the $\chi^2$ will be used to project out modes for the auto-correlation power spectrum. On average, LSS data such as quasar and galaxy surveys should not be correlated with observational systematics, \eg dust extinction, seeing variation or stellar density. Individual realisations may exhibit non-zero cross-spectra, but, in the absence of systematics, these should be consistent with zero within the statistical uncertainties resulting from the estimator and the noise. Therefore, the cross-spectra can be used for null tests, yielding one reduced $\chi^2$ per systematics mode, to decide if the mode should be projected out. Large $\chi^2$ indicates significant spurious correlation between the data and the mode, pointing to contamination by systematics. For a given set of uncorrelated systematics, one can impose a global $\chi^2_{\rm cut}$ and only project out the modes with null test $\chi^2$ above this value. The $\chi^2_{\rm cut}$ can be fixed according to absolute criteria, for example $\chi^2_{\rm cut}=1.0$, which ensures that the remaining (\ie non-projected) modes pass the null tests and are consistent with zero within the statistical uncertainties. Alternatively, one can also adjust $\chi^2_{\rm cut}$ to control the number of projected modes --- and thus, the increase in variance --- in the estimated power spectra. The next section provides more details about the power spectrum estimator used for the null tests.

In conclusion, extended mode projection only relies on the ability to create systematics templates, decorrelate them, and cross-correlate them with the data of interest. It allows one to consider generic contamination models with large numbers and complex combinations of templates, from which the main orthogonal modes will be extracted. Under the reasonable assumption that these systematics --- and thus the orthogonal modes --- do not correlate on average with the data, one can efficiently find the modes which need to be projected out.  Compared to other techniques previously applied to galaxy and quasar survey data  \citep{ross2011weights, ross2012systematics, agarwalho2013xdqsoz, hoagarwal2013xdqsoz}, extended mode projection does not attempt to directly correct the data or the power spectra but rather, analytically marginalises over the systematics when computing band-power estimates. Beyond the parameters of the cross-power spectrum estimator (\eg band-power width), which are guided by the sky coverage and noise level, the only tunable parameters in this technique are the set of input systematics templates and $\chi^2_{\rm cut}$, which corresponds to the accepted level of correlation between a potential systematic and the data. Therefore, this framework is based on the principles of blind analysis, and can be used in a variety of situations where the data are non-linearly contaminated by several systematics.

\subsection{Fast cross-power spectra and null-tests} \label{sec:chi2approx}

With our implementation of extended mode projection, the systematics modes to be projected out are selected based on null tests resulting from cross-power spectra with the data of interest. Due to the potentially large number of modes, it is essential to use a fast estimator to compute these cross-power spectra. The PCL estimator is a good candidate, but may be suboptimal when using complex masks, or if the cross spectra are significantly non-flat, which is likely when considering significantly contaminated data such as quasar surveys \citep{Leistedt2013excessdr6}. Yet, using the QML estimator would require significant time and memory resources, despite these power spectra only being used for null tests, and thus not requiring the same level of accuracy as the main $\mCl$ analysis.  Therefore, we designed an approximate QML estimator which yields fast null tests which are quasi-optimal when compared with the full QML estimator. This new estimator simply replaces the Fisher matrix in $\mat{E}^b_{\rm QML}$ by  the inverse covariance matrix of the PCL estimator. More precisely, we compute cross spectra between one data sample $\vec{x}_{\rm map}$ and a systematics mode $\vec{x}_{\rm sys}$ using
\equ{
	 \hat{C}^{\rm null test}_b = \sum_{b'} \mat{G}^{-1}_{bb'} \ \frac{1}{2} [\vec{x}_{\rm map}^t \mat{C}_{\rm map}^{-1} {\mat{P}^{b'}} \mat{C}_{\rm sys}^{-1} \vec{x}_{\rm sys}], \label{approxestimator}
}
where $\mat{G}_{b_1b_2}$ is the covariance of the PCL estimator, which can be simplified to
\eqn{
	\mat{G}_{b_1b_2} &=&  \sum_{b_1 b_2 b_3 b_4} \mat{M}^{-1}_{b_1 b_3}  \mat{M}^{-1}_{b_2 b_4} C^{\rm map}_{b_3} C^{\rm sys}_{b_4} 4 \boldsymbol{\Pi}_{b_1 b_2 b_3 b_4} ,
}
where $ \mat{M}$ is the PCL coupling matrix. Here, $\boldsymbol{\Pi}_{b_1 b_2 b_3 b_4} = \tr [\mat{P}^{b_1}\mat{P}^{b_2}\mat{P}^{b_3}\mat{P}^{b_4}]$ is computationally intensive to evaluate but only depends on the mask under consideration, and can therefore be precalculated. In this formulation, $C^{\rm map}_{b}$ and $C^{\rm sys}_{b}$ are priors for the band-powers of the data and the systematics mode to be cross correlated, also used to construct the covariance matrices $\mat{C}_{\rm map}$ and $\mat{C}_{\rm sys}$. In particular, $C^{\rm map}_{b}$ is the theoretical prior also used in the main optimal estimator, whereas we obtain $C^{\rm sys}_{b}$ using a fast PCL estimator applied to $\vec{x}_{\rm sys}$. As a consequence, the approximate power spectrum estimator of \equref{approxestimator} is fast and can be used to efficiently calculate the numerous cross-power spectra between the data and the systematics. In this work, we use null test $\chi^2$s which are simply calculated using a Gaussian likelihood comparing the $\hat{C}^{\rm null test}_b$ with zero, as in \cite{Leistedt2013excessdr6}. In a later section, we compare the $\chi^2$ values obtained by the optimal and the approximate power spectrum estimators, and show that the latter is sufficiently accurate for the masks, data and systematics under consideration.

\begin{figure}
\includegraphics[trim = 0.0cm 0.0cm 0.0cm 0.0cm, clip, width=8.4cm]{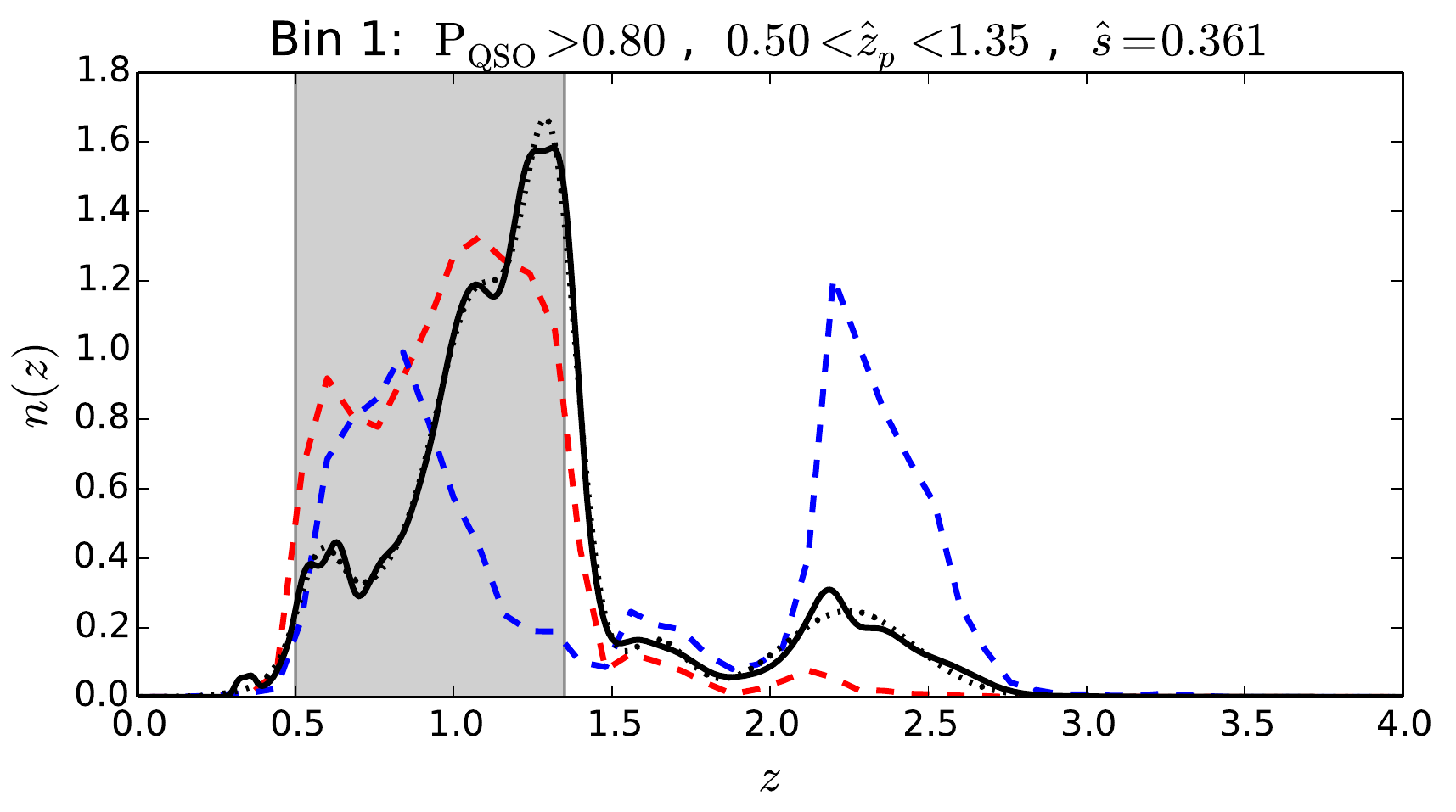}\\
\includegraphics[trim = 0.0cm 0.0cm 0.0cm 0.0cm, clip, width=8.4cm]{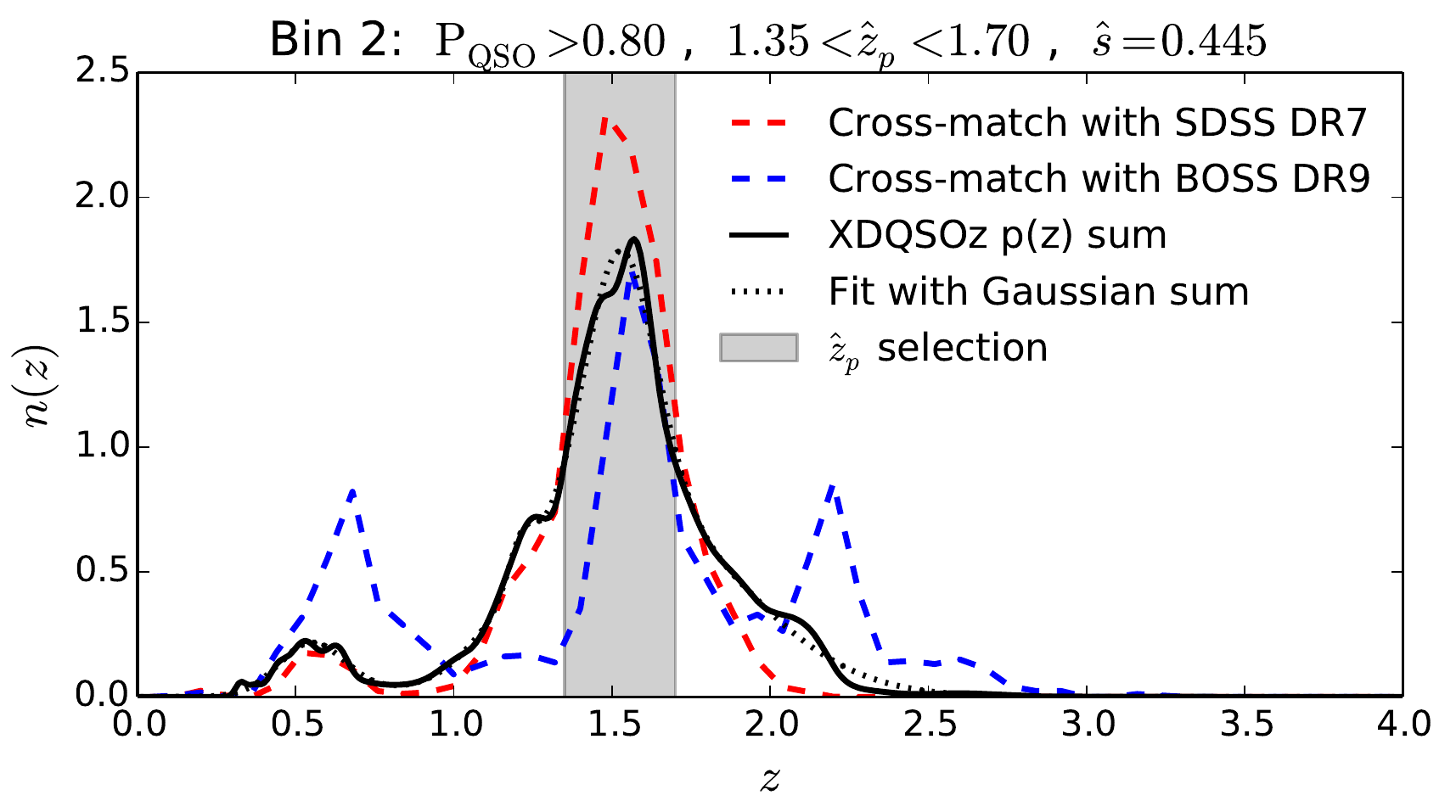}\\
\includegraphics[trim = 0.0cm 0.0cm 0.0cm 0.0cm, clip, width=8.4cm]{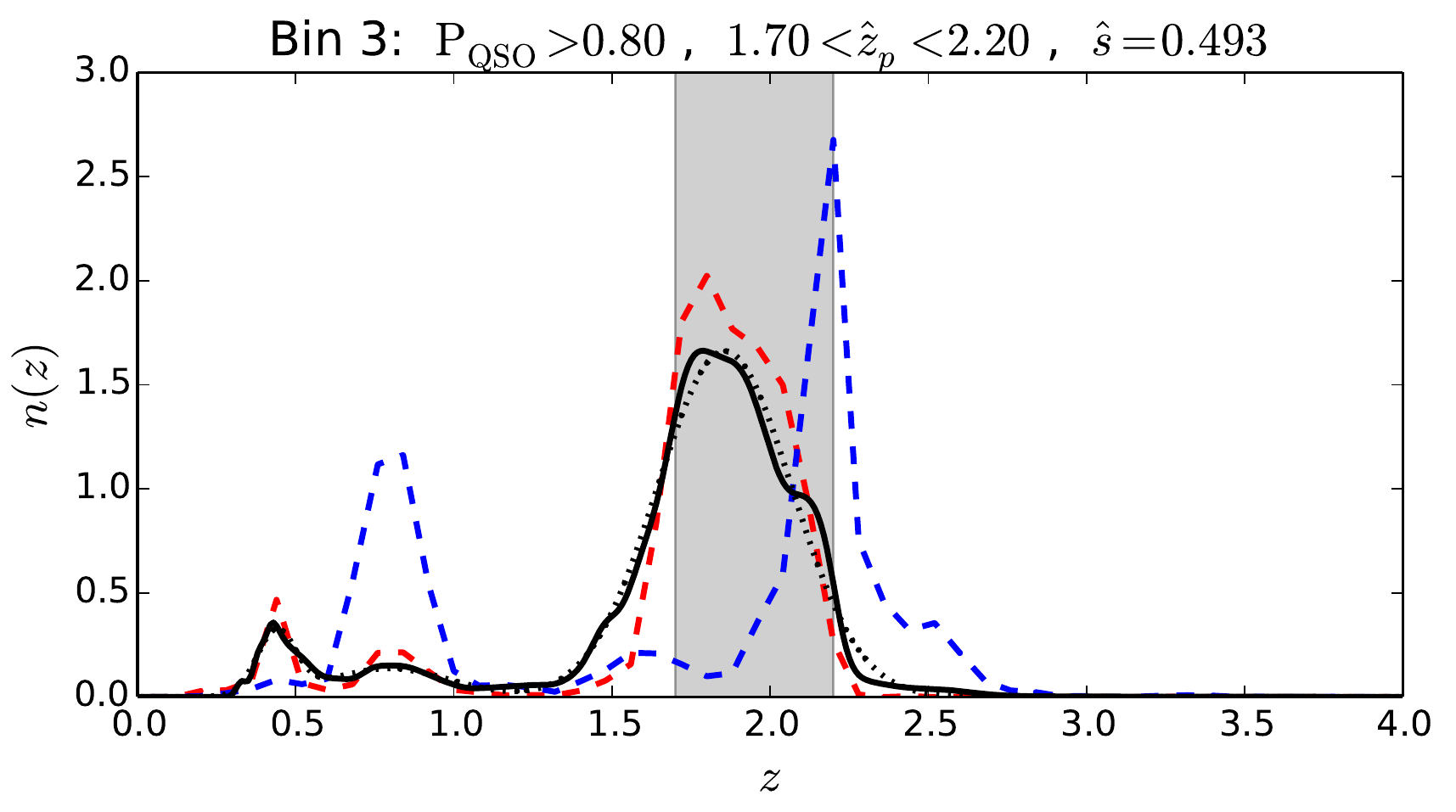}\\
\includegraphics[trim = 0.0cm 0.0cm 0.0cm 0.0cm, clip, width=8.4cm]{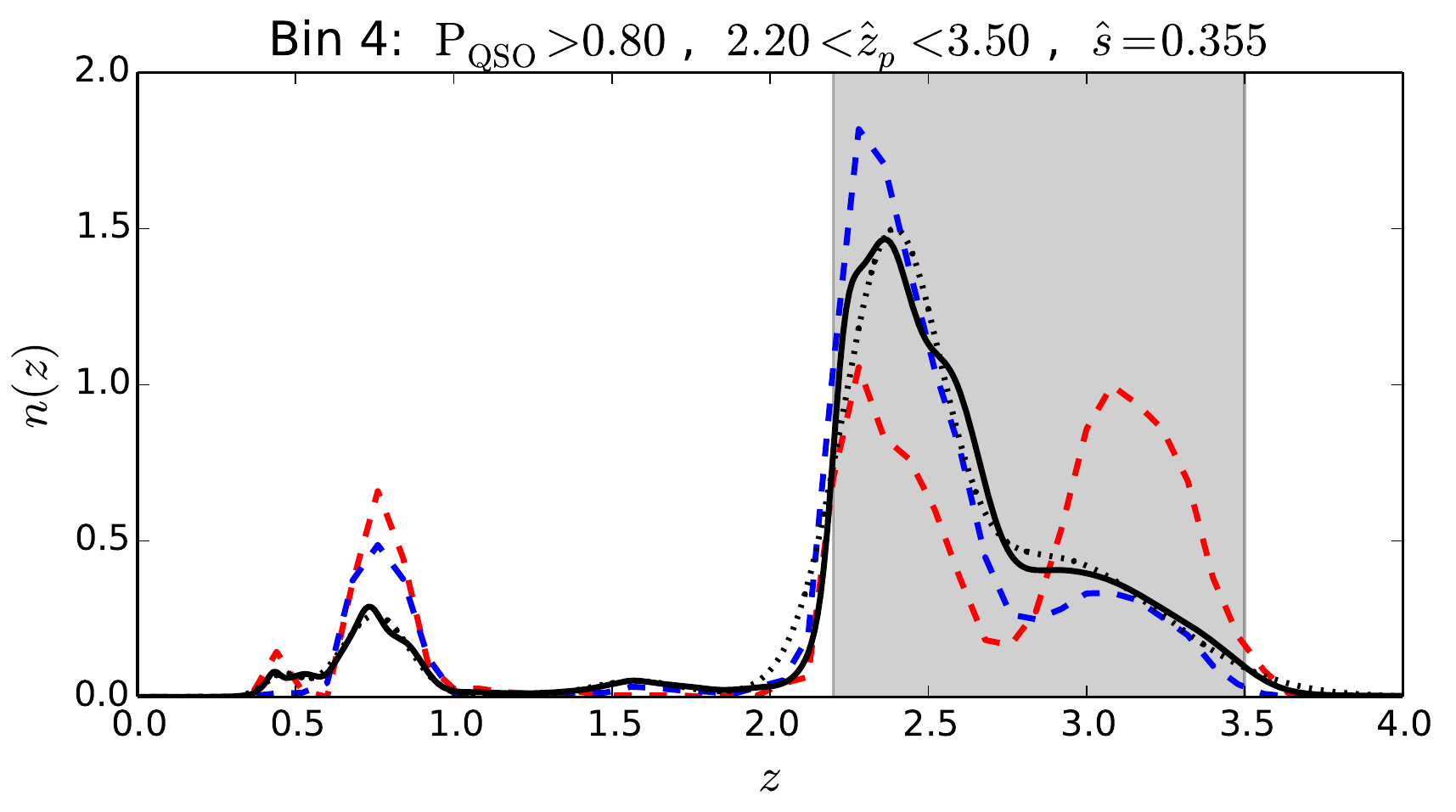}
\caption{Redshift distributions of the photometric quasar samples selected with $\pqso>0.8$ and top hat windows on the photometric redshift estimates $\hat{z}_p$, highlighted by the grey regions. The black curves show redshift distribution estimates $n(z)$ of the underlying quasar distributions, calculated as the sum of the posterior distributions on the individual redshift estimates. The dashed line shows a fit with Gaussian kernels, used for calculating the theoretical $C_\ell$'s. The red and blue dashed curves are redshift histograms of the objects cross-matched to the SDSS DR7 and BOSS DR9 spectroscopic quasar catalogues. These are only used for qualitative purposes, since they provide estimates of $n(z)$ which are biased towards low- and high-redshift, respectively, due to the magnitude and redshift selection effects (details in the text).}
\label{fig:redshiftdistributions}
\end{figure}

\section{Application to SDSS photometric quasars}

We now apply the extended mode projection framework to photometric quasars from the SDSS, and compare the resulting power spectrum measurements with theoretical predictions. We discuss these measurements, their suitability for cosmological inference, and the ability of extended mode projection to eliminate spurious correlations due to spatially-varying systematics. 

\subsection{Sample selection and redshift distributions}

We consider the XDQSOz catalogue of photometric quasars \citep{bovy2012xdqsoz}, which contains quasar candidates selected from the set of point sources of the SDSS DR8 imaging data, covering $\sim 10^4$ deg$^2$ of the southern and northern Galactic sky. For each object, XDQSOz includes the probability of being a quasar or a star, calculated from the observed magnitudes ({\it ugriz}) and their estimated errors, using a data-driven model of the density distribution of quasars in {\it ugriz} flux-redshift space. This model was constructed by applying the {\it extreme deconvolution} algorithm \citep{Bovy2011xd} to the spectroscopically-confirmed quasars in the SDSS DR7 quasar catalogue \citep{Schneider2010qsodr7cat}. XDQSOz also includes photometric redshift estimates $\hat{z}_p$, defined as the highest peak in the posterior distribution of the individual photo-$z$ estimates. Although the latter distributions are not directly released in the public version of XDQSOz, they can be straightforwardly recalculated using the flux-redshift space model and the publicly-available code. 

We consider all objects with $\pqso>0.8$, but with no further cuts to the data, other than the sky masks described in the next section. We separate this catalogue into four samples by selecting objects with photometric redshifts $\hat{z}_p$ in top-hat windows ranging $[0.5,1.35]$, $[1.35,1.7]$, $[1.7,2.2]$ and $[2.2,3.5]$, which have comparable numbers of objects, and thus similar shot noise. We reject the low-redshift objects ($\tilde{z}_p < 0.5$);  this redshift range suffers from nontrivial incompleteness issues due to missed low-$z$ quasars, which are resolved and therefore not processed by the point source star-quasar classifier.

\begin{figure*}
\vbox to220mm{\vfil\hspace*{-3mm}\includegraphics[width=17cm]{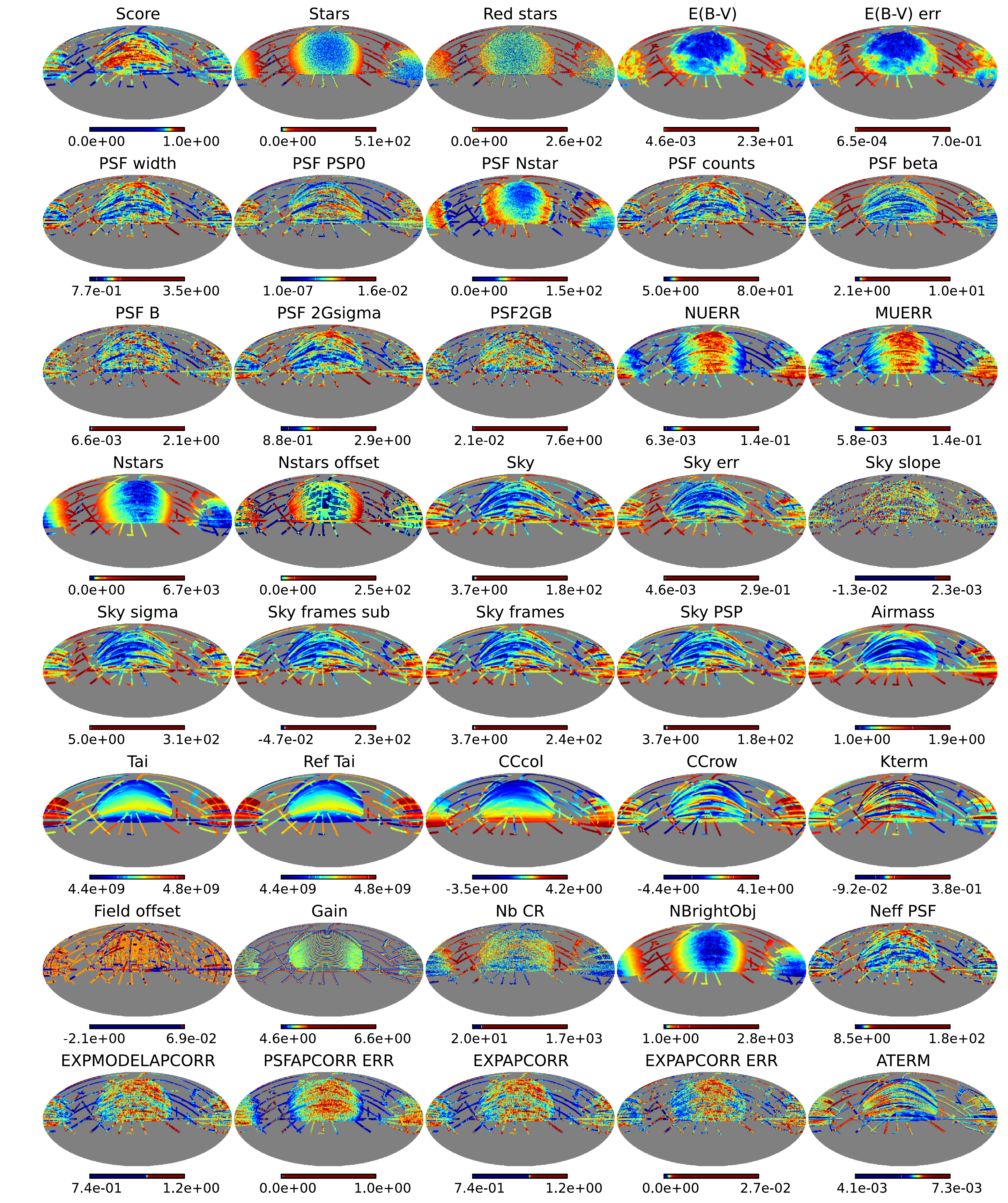}
\caption{Templates of the main systematics selected as potential contaminants in the photometric quasar catalogues, and used within the extended mode projection approach. Apart from the first row, all systematics are mapped in the {\it ugriz} bands and included in the analysis, but only templates for the {\it i} band are shown here. More details can be found in the text, and the full descriptions of the nature and units of these quantities (only referred to by their abbreviated names in this paper) can be found in the SDSS database and documentation.} \label{fig:systematics}\vfil}
\end{figure*}

\begin{figure*}\label{fig:uncorrsystematics}
\hspace*{-3mm}\includegraphics[width=17cm]{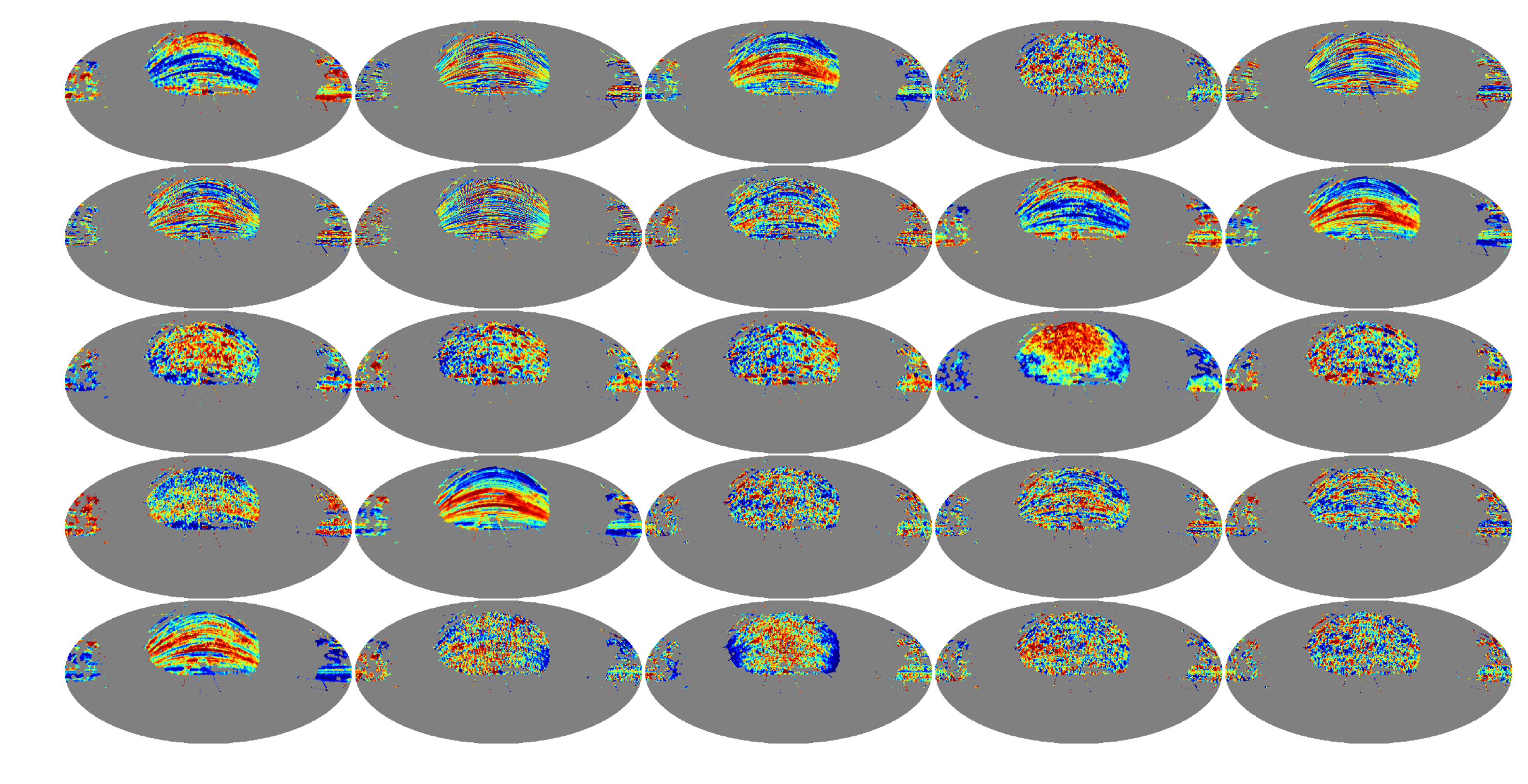}
\caption{A subset of systematics templates originating from the decorrelation (using Mask 1) of the basic set of systematics, a subset of which is presented in \figref{fig:systematics}.}
\end{figure*}

In order to calculate theoretical predictions for the angular power spectra of the four samples, we follow \cite{Leistedt2013excessdr6} and use \textsc{camb\_sources} \citep{challinorlewis2011cambsources} to compute and project the 3D matter power spectrum into angular auto- and cross-power spectra. We fix the cosmological parameters to \textit{Planck} $\Lambda$CDM best-fit values\footnote{Specifically, $\Omega_ch^2=0.118, \Omega_m=0.305, \Omega_bh^2 = 0.0221 , H_0=67.7\ {\rm km s}^{-1} {\rm Mpc}^{-1}, \ln(10^{10}A_s)=3.09, n_s=0.961, \tau = 0.095$, ${\rm N}_{\rm eff}=3.046$, {\rm\ and \ } $\sum m_{\nu}=0.05$.} \citep{Planck2013cosmologicalparams}. Importantly, although the matter power spectrum only depends on cosmological parameters, the projected power spectra require additional information about the samples under consideration: the unit-normalised redshift distribution of tracers $n(z)$, and the logarithmic slope of the number counts $s$, to account for the effect of magnification due to lensing. We estimated $s$ for the four samples by calculating the slope of the histogram of number counts in terms of the $g$-band PSF magnitude at $g = 21$. We found $\hat{s}=0.361, 0.445, 0.493$, and $0.355$, respectively, for these samples. 

The redshift distributions are nontrivial to estimate because photometric redshift estimates of quasars suffer from large uncertainties (typically $\sigma(\hat{z}_{\rm phot}) \sim 0.3$) and large fractions of outliers (see \eg \citealt{Richards2008rqcat, bovy2012xdqsoz}). In the present case, the actual $n(z)$ of the quasar samples extend significantly beyond the windows used to construct them. This not only introduces uncertainties in the theoretical predictions, but also cosmological information in the cross-power spectra, since the $n(z)$ of the samples are likely to overlap with each other, as detailed in the next sections. One solution to reduce the associated uncertainties is to employ low-resolution histograms, which are more robust to individual \photoz\ uncertainties, but do not capture all the structure of the underlying $n(z)$. Alternative methods also exist to attempt to estimate high-resolution $n(z)$ from the poorly-determined photometric redshifts (\eg \citealt{Matthews:2010an, McQuinn:2013ib}). In \cite{Leistedt2013excessdr6}, we used a cross-matching technique to obtain unbiased estimates for the $n(z)$ of the \cite{Richards2008rqcat} catalogue SDSS DR6 photometric quasars. In our case the XDQSOz catalogue includes posterior distributions for the redshift estimates of individual objects, which are computed with the same data-driven model used for the classification, and trained with the best spectroscopic data available. Hence, a simple estimator for $n(z)$ is the sum of the posterior distributions of the individual objects in each sample. \figref{fig:redshiftdistributions} shows the resulting $n(z)$, as well as a fit using a superposition of Gaussian kernels for use in \textsc{camb\_sources}. We also show the redshift histograms of cross-matched catalogues of quasars found in catalogues of spectroscopically-confirmed quasars from SDSS DR7 and BOSS DR9 \citep{Schneider2010qsodr7cat, paris2012bossqsodr9}. We applied the relevant flags and cuts to create spatially-uniform versions of these catalogues, \ie with constant depth, in order to avoid spurious selection effects due to depth variations. However, SDSS DR7 mostly contains bright objects, and BOSS DR9 is dominated by $z>2.2$ quasars. Therefore the dashed redshift histograms shown in \figref{fig:redshiftdistributions} are biased estimates of $n(z)$, but nevertheless confirm the features found in the XDQSOz $n(z)$. In particular, all four samples contain low-redshift quasars ($z\sim0.7-1.0$), mainly because of their large scatter in {\it ugriz} colour space which overlaps with that of higher redshift quasars -- hence the difficulty of separating them \citep{Bovy2010xdqso}. The corrections required to follow \cite{Leistedt2013excessdr6} and obtain unbiased $n(z)$ estimates through cross-matching are more difficult to calculate for XDQSOz than the \cite{Richards2008rqcat} catalogue, because the former extends to fainter magnitudes and higher redshifts. In particular, using the full spectroscopic catalogue requires spatially-dependent corrections, which are much more uncertain due to sample variance. For these reasons, we did not attempt to derive and apply these corrections, but rather used the $n(z)$ obtained using the stacked XDQSOz redshift posterior distributions. This should provide the most robust solution, since the XD data-driven model has not been demonstrated to suffer from any biases or systematics over the redshift ranges of interest. 

In this paper, we focus on evaluating the quality of the measurements and the capacity of mode projection to mitigate systematics. Hence, we will fix the parameters used to calculate the theoretical angular power spectra, and also observational parameters such as the shot noise and the bias of quasars. The latter is assumed to be scale-independent and fixed to 
\equ{
	b(z) = \left[ 1 + \left(\frac{1+z}{2.5}\right)^5 \right],
}
which is consistent with observations (\eg \citealt{porcianiNorberg2006, SlosarHirata2008, Sherwin2012qsolensing}). As shown in subsequent sections, these parameters provide a good fit to the measured power spectra. However, it is essential to consider their uncertainties when testing models and inferring cosmological parameters. Hence, all theoretical and observational parameters will be varied in the companion paper \citep{LeistedtPeiris:2014:XDQSO_png}, where we constrain PNG and evaluate the robustness of these constraints to the underlying models and assumptions.

\subsection{Masks and systematic uncertainties}

The main source of systematics in photometric quasar catalogues is stellar contamination, which arises because separating stars from quasars using imaging data is a difficult task. With the $\pqso>0.8$ cut, one can expect a significant fraction of the selected photometric quasars to be stars. However this proportion further depends on the redshift range of interest, since the colours of quasars, and thus the separability of the stellar and quasar loci, evolve with redshift. In addition, observing conditions unavoidably vary with time, and thus on the sky, as data are acquired. This affects both the number of detected point sources and the star-quasar separation, thus creating spatially-varying depth and stellar contamination. We have mapped as many of the potential systematics as possible in order to create templates for the mode projection framework. In what follows, we detail the construction of a base set and an extended set of systematics templates, which will be projected out when estimating the quasar power spectra. 

\begin{figure}
\includegraphics[trim = 0.0cm 1.8cm 0.0cm 1.0cm, clip, width=8cm]{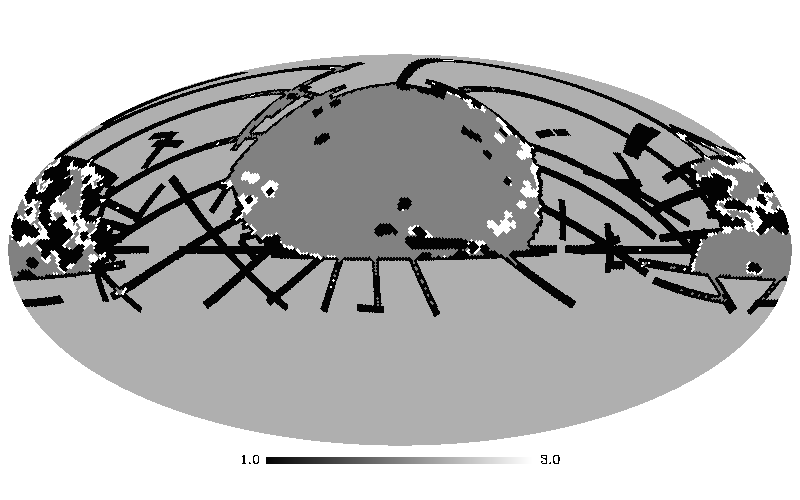}
\caption{Masks constructed for the power spectrum analysis of the XDQSOz samples. Mask 1 (grey + white) and Mask 2 (grey only) are created by applying dust extinction cuts $E(B-V)>0.1$ and $E(B-V)>0.08$ to the full SDSS DR8 coverage (black region), in addition to the common cuts {\sffamily Psfwidth}$_i>2.0$ and {\sffamily Score}$>0.6$ to avoid bad quality regions.  }
\label{fig:masks}
\end{figure}

For Galactic dust extinction, which affects the properties of point sources and photometric colours, we used the maps of $E(B-V)$ from \cite{Schlegel1998dust} and {\it Planck} \citep{Planck2003dust}. They differ in some regions of the sky, especially near the Galactic plane, but are correlated, so we include both maps in the set of systematics prepared for extended mode projection. For the stellar density, we constructed a stellar density map from the SDSS DR6 point sources selected with $18.0 < r <18.5$ and $i < 21.3$, and a second map with an additional cut $g-r >1.4$ to select red stars \citep{PullenHirata2012}. We retrieved the data for the calibration and observing conditions from the \textsc{Fields} table in the SDSS CAS server. We mapped all quantities on the sky directly on the \textsc{healpix} \citep{healpix1} grid, as the quasar maps are also manipulated in this format. This solution does not use the exact geometry of the SDSS tiles (which would require the use of the \textsc{mangle} software \citep{Hamilton2004mangle, Swanson2008mangle}) but is much faster while being sufficiently accurate on the scales of interest. \figref{fig:systematics} shows the base set of systematics we include in this study. Apart from the first row, all templates were constructed for the five \textit{ugriz} bands, and only the \textit{i} band is shown in \figref{fig:systematics}. Also, the following quantities were not included on the figure for space reasons: {\sffamily ra, dec, run, fields, mjd, mustart, nustart, total, devapcorrection, devapcorrectionerr, devmodelapcorrection, devmodelapcorrectionerr, nfaintobj, nstaroffset}. Full details of the nature, description and units of these quantities can be found in the SDSS database and documentation.

The total number of templates in this set of base systematics is 220, but after decorrelating these templates only $\sim 100$ have significant eigenvalues, the others being consistent with numerical noise, \ie redundant modes in the system of templates. A subset of these orthogonal modes is shown in Fig.~3. With such a small number of modes, all can be projected out when estimating power spectra, since they only generate a small increase in variance $\mat{V}_{bb'}$. However, this setting will only treat the spurious clustering due to linear combinations of these systematics. The actual contamination signal is likely to be non-linear and therefore may not be effectively eliminated. To consider non-linear contamination model, we create an extended set of templates, also including products of pairs of the base systematics. Once decorrelated, the $\sim22,000$ maps were reduced to $\sim3,700$ modes. Since they are orthogonal modes, projecting out all of them would be equivalent to removing $\sim3,700$ data points in some rotated basis, thus significantly increasing the variance of the estimates. For this reason, we now use the extended mode projection approach to project out only those modes which are significantly correlated with the quasar data, as detailed in the next section. 

Although extended mode projection considers non-linear contamination models, it is unlikely to fully clean the stellar contamination across the whole survey area, in particular in the worst regions where systematics are significant. Hence, we construct two masks, shown in \figref{fig:masks}. Both exclude pixels with $\textsc{Psfwidth}_i>2.0$ and $\textsc{Score}>0.6$. In addition, Mask 1 uses $E(B-V)>0.1$ and Mask 2 uses $E(B-V)>0.08$, to remove more dusty regions in the South Galactic Cap. The masks were extended to avoid smoothing-induced contamination, as explained in \cite{FPP11} and \cite{Leistedt2013excessdr6}. Note that previous SDSS quasar studies used more aggressive masks \citep{hoagarwal2013xdqsoz, PullenHirata2012, Leistedt2013excessdr6}, and the point of extended mode projection precisely is to avoid extra sky cuts and obtain finer control over the elimination of contamination by projecting out the relevant systematics.

\subsection{Estimation settings and blind mitigation of systematics}

We used the QML estimator to simultaneously measure the auto- and cross- angular power spectra of the four XDQSOz samples. We used \textsc{healpix} resolution of $\nside=64$ for all quasar maps, masks and systematics templates in this work, in which case the power spectrum can be estimated accurately up to $\ell_{\rm max}\sim 130$, which is a satisfactory band-limit given the noise levels and the sky cuts under consideration, and the computational requirements of the estimator. The choice of the multipole bin size is guided by the variance of the estimates. For most cases we adopted a bin size $\Delta\ell = 21$ to obtain a good compromise between variance and multipole resolution of the estimated power spectra. We also used $\Delta\ell = 15$ for the final runs in order to obtain a better resolution on the lowest multipoles, and therefore a better sensitivity to the PNG signal. The resulting constraints on PNG are presented in the companion paper \citep{LeistedtPeiris:2014:XDQSO_png}, and are indeed more stringent in that setting. For the fiducial power spectrum priors required for QML, we used the theoretical power spectra detailed in the previous sections. The QML estimator is robust to small changes in these values, especially when measuring the auto- and cross- spectra, which are relatively featureless\footnote{Since quasars span large volumes, the projection integrals weaken features such as the baryon acoustic oscillations. The latter would only be detectable using narrow redshift samples, which cannot be done with XDQSOz due to the large uncertainties of the photometric redshift estimates.}, and simultaneously estimated from the four quasar maps and the full prior covariance matrix.

\begin{figure}
\centering
\includegraphics[width=8cm]{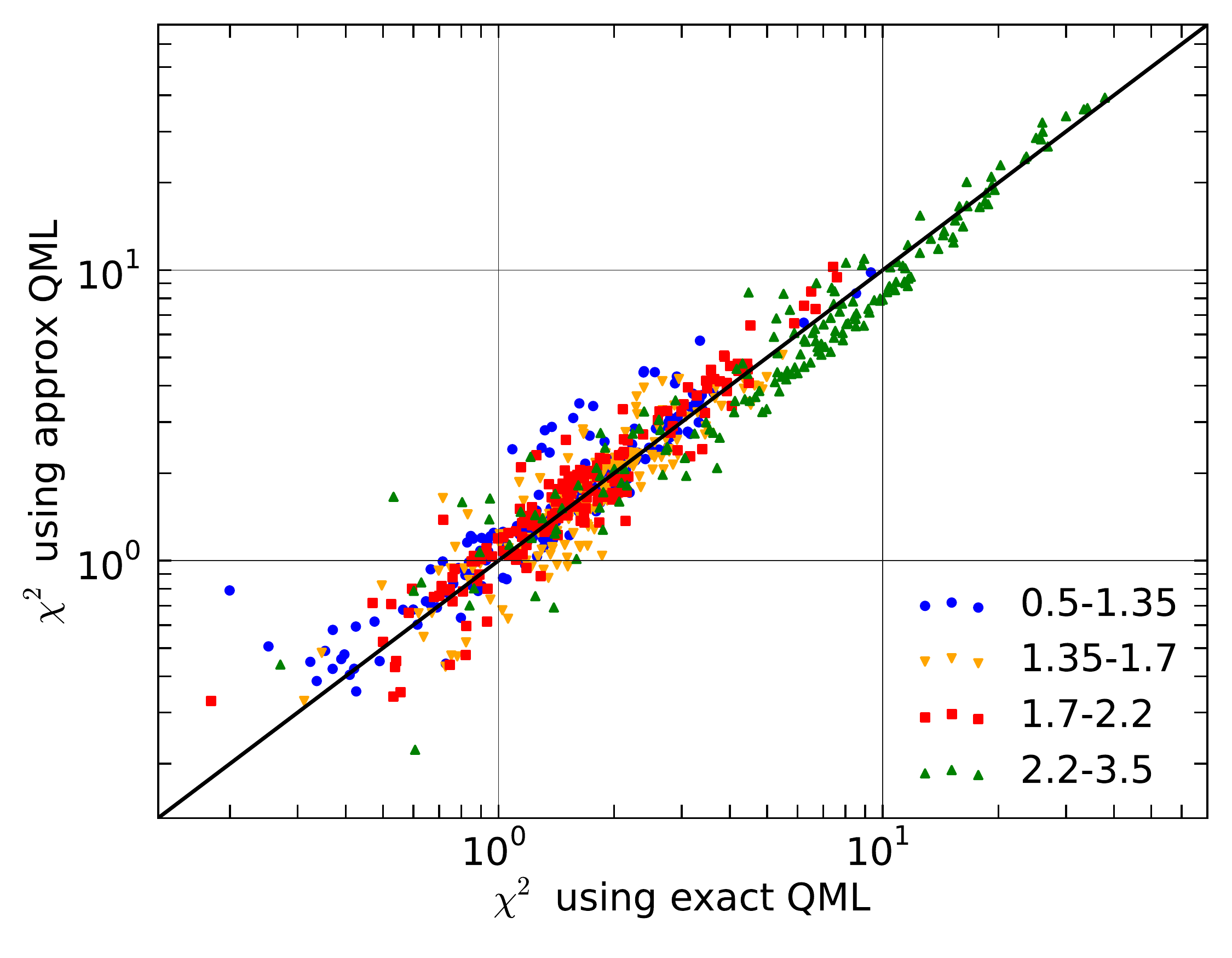}
\caption{Comparison of the reduced $\chi^2$ obtained with the optimal and approximate QML cross-correlation estimators. Each dot is a value obtained by using both estimators to calculate the cross-spectra between the four quasar samples (colour-coded) and the base set of $220$ systematics templates.}
\label{fig:chi2comparison}
\end{figure}

\begin{figure}
\includegraphics[trim = 0.4cm 0.4cm 0.4cm 0.4cm, clip, width=8cm]{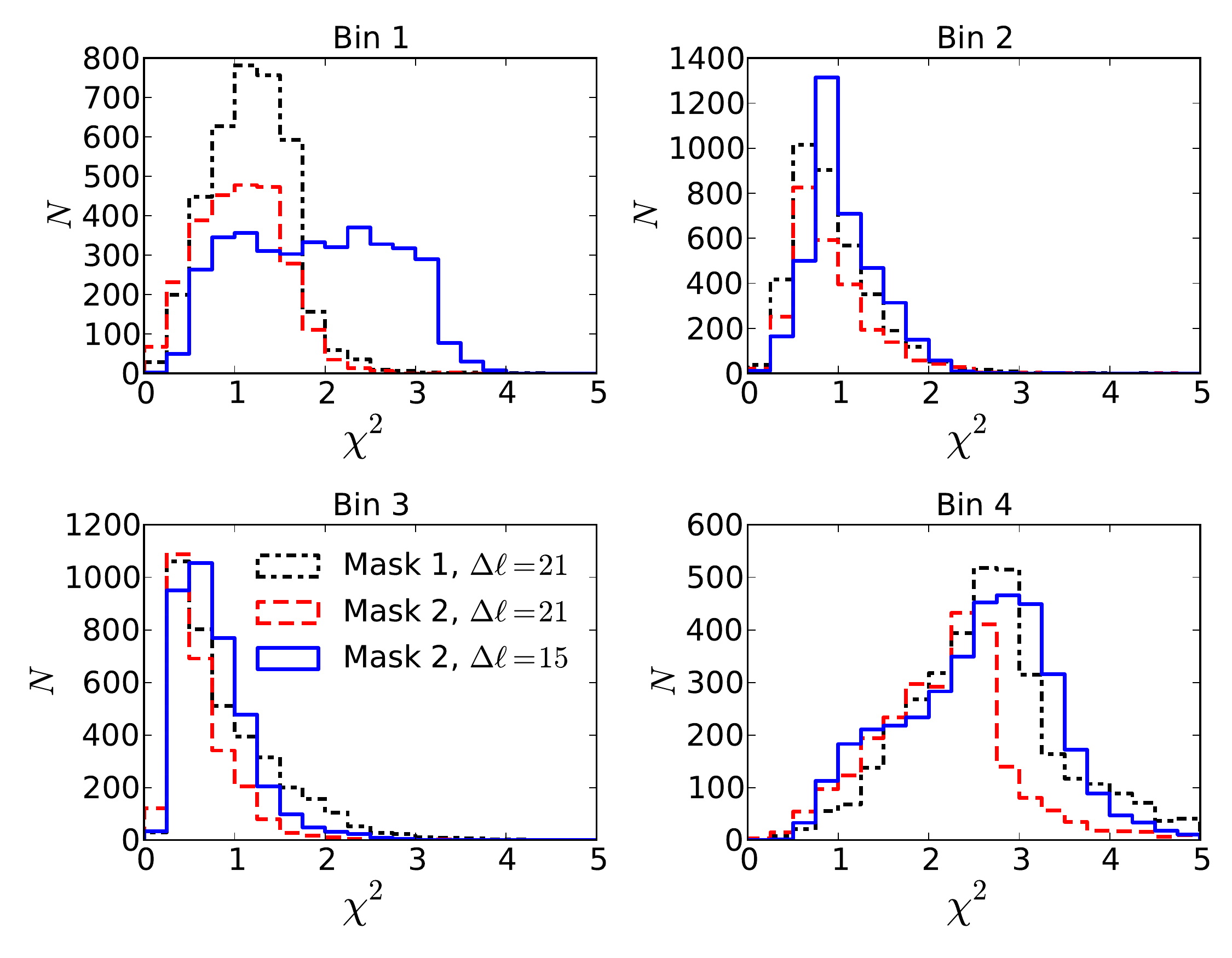}
\caption{Distributions of the reduced $\chi^2$ arising from null tests performed with the cross-power spectra of the four quasar samples (four panels) with the $\sim 3700$ orthogonal modes of the extended set of systematics templates ($\sim 200$ base templates plus products of pairs). The last step of extended mode projection consists of choosing a $\chi^2_{\rm cut}$ to decide which modes will be projected out when estimating the angular power spectra of the quasar samples.}
\label{fig:chi2distribs}
\end{figure}

For the power spectrum measurements presented in the next section, we considered three settings for the treatment of systematics (in combination with the two masks of \figref{fig:masks}): no mode projection (``no mp"), projection of the base set of systematics ``basic mp", $\sim100$ uncorrelated modes), and projection of the extended set plus products of pairs of templates (``ext. mp", $\sim3700$ uncorrelated modes). For the latter, we followed our implementation of { extended mode projection}, and only projected the modes which were significantly correlated with the data. To perform this selection, we cross-correlated the quasar maps with all orthogonal modes using the approximate QML estimator presented in Section~\ref{sec:chi2approx}, using the same setting as the main estimator ($\nside=64$,  $\ell_{\rm max}=130$, $\Delta\ell = 21$ or $\Delta\ell = 15$). The resulting cross-spectra were used as null tests, and we calculated a reduced $\chi^2$ per systematics mode using a simple Gaussian likelihood \citep{Leistedt2013excessdr6}. \figref{fig:chi2comparison} shows the comparison of the reduced $\chi^2$ obtained with the optimal and approximate QML estimators detailed in Section~\ref{sec:chi2approx}, and demonstrates that the latter is sufficiently accurate for the purpose of these null tests. We observe no significant bias, a reasonable scatter around the axis $\chi^2_{\rm opt. QML} = \chi^2_{\rm approx. QML}$, and a small number of outliers. In particular, we see that selecting modes for mode projection with the cut $\chi^2_{\rm approx. QML} > 1$ reproduces the selection using a cut $\chi^2_{\rm opt. QML}>1$ reasonably well, validating the use of the approximate estimator in the null tests.    

\begin{figure*}
\includegraphics[width=18cm]{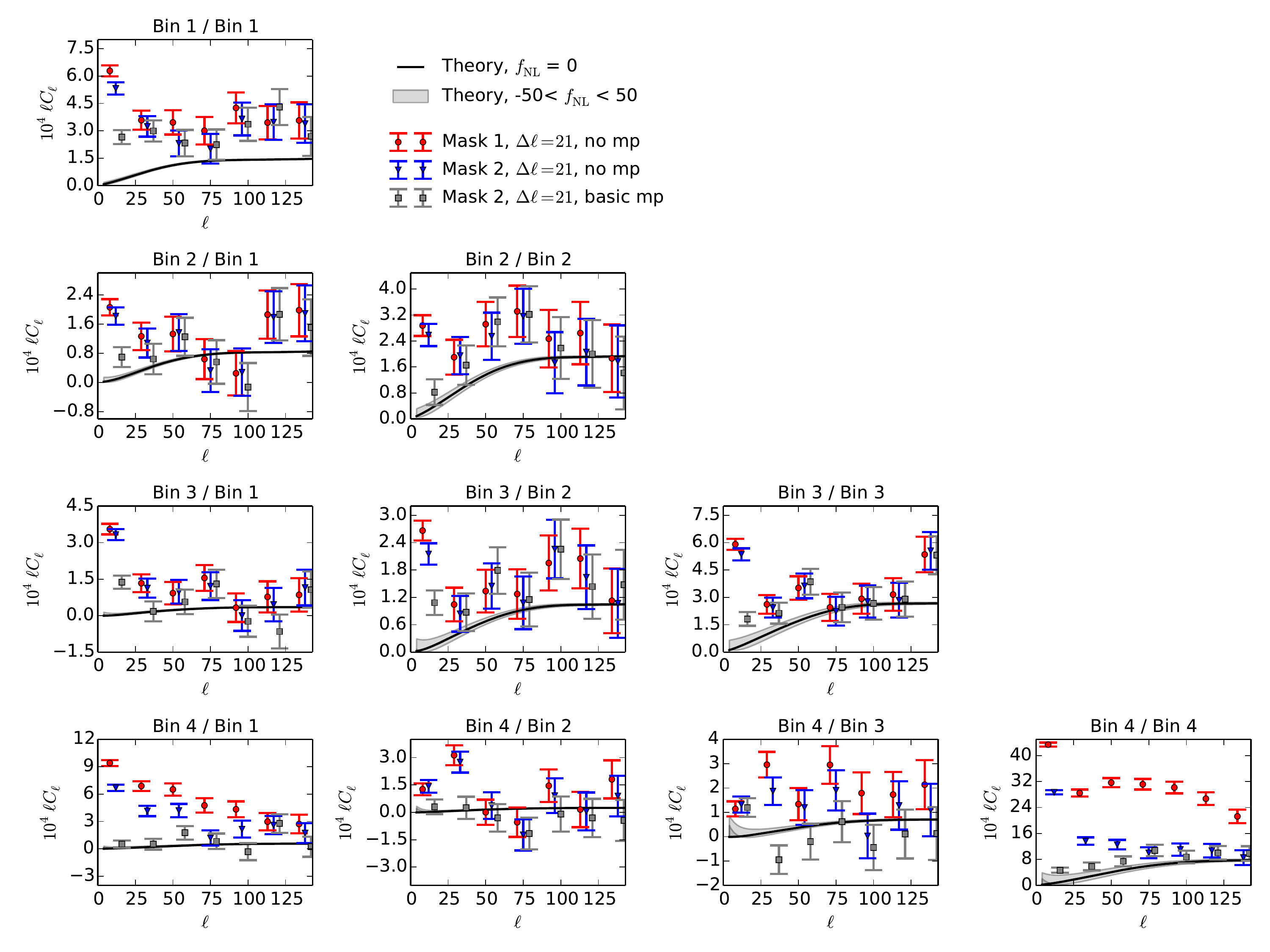}
\caption{Auto- and cross-angular power spectra of the four XDQSOz quasar samples measured with the quadratic maximum likelihood (QML) estimator, for the two masks in \figref{fig:masks}, without (``no mp") and with (``basic mp") mode projection of the base set of systematics ($220$ templates from the SDSS \textsc{Fields} table, shown in \figref{fig:systematics}). The solid lines show our fiducial theoretical power spectra from a {\it Planck} best-fit cosmology and $\fnl=0$, and the shaded bands show the excursion region allowed when varying $\fnl$ in $[-50,50]$ (see text for details).  }
\label{fig:resultsnomp}
\end{figure*}

\begin{figure*}
\includegraphics[width=18cm]{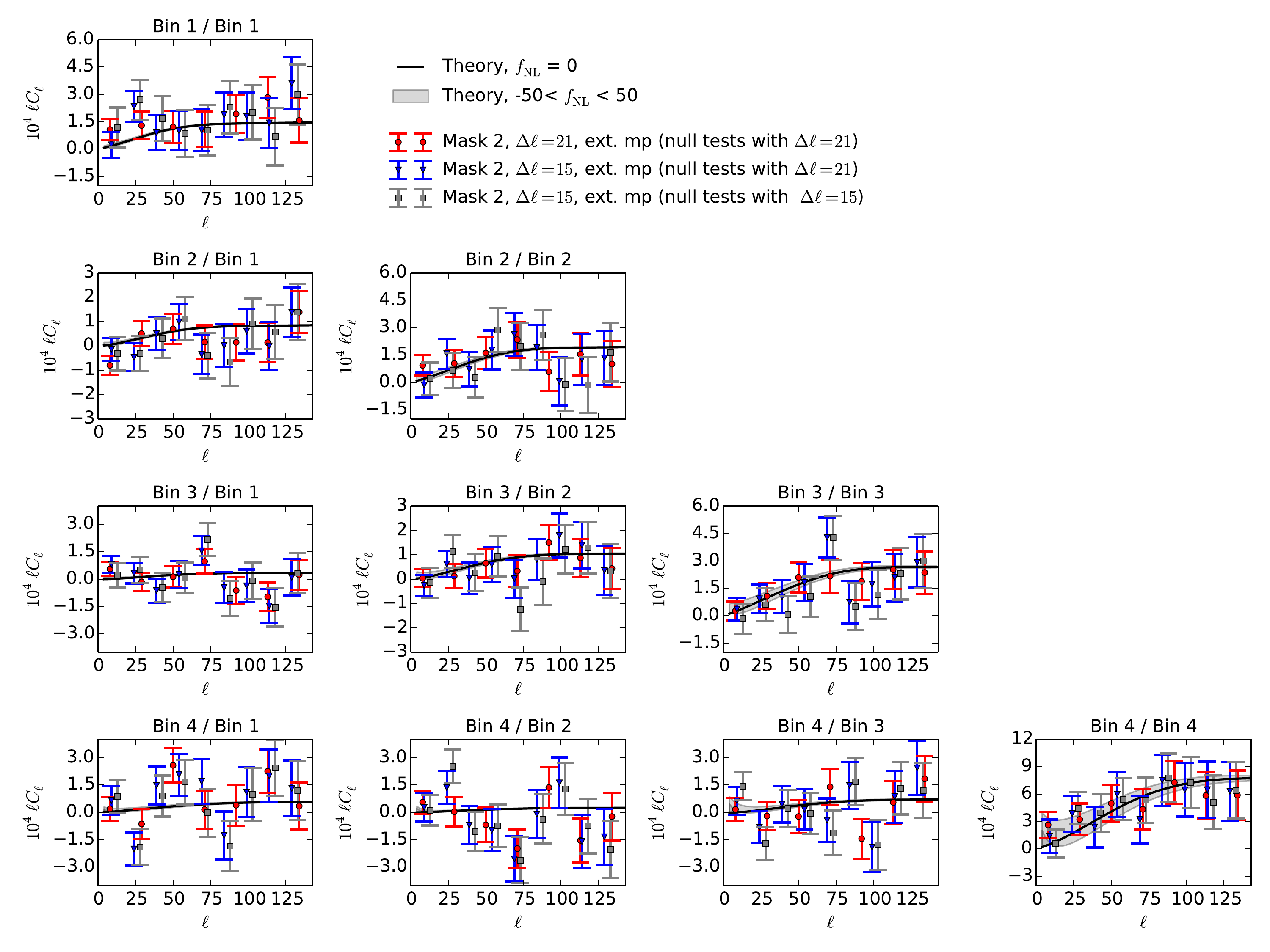}
\caption{Same as \figref{fig:resultsnomp}, but using extended mode projection. We used a non-linear contamination model incorporating the base set of systematics templates and also pairs of products of templates, yielding $\sim 22,000$ templates: but only $\sim 3,700$ orthogonal spatial modes remained after decorrelation. We calculated QML cross-angular power spectra to perform null-tests and calculated reduced $\chi^2$, as shown in \figref{fig:chi2distribs}.}
\label{fig:resultsemp}
\end{figure*}

\figref{fig:chi2distribs} shows the distributions of the reduced $\chi^2$ values obtained by cross-correlating the four quasar maps with the $\sim3700$ modes. Interestingly, these distributions can be used to qualitatively evaluate the level of contamination of the samples, and also the effect of masking. In particular, the second and third samples are the least contaminated, and the $\chi^2$ distributions are not improved by using the second mask instead of the first mask, in addition to being robust to the multipole resolution ($\Delta\ell = 21$ or $\Delta\ell = 15$). The null tests also demonstrate that the first and fourth samples are significantly contaminated by systematics, since a large fraction of the modes have large reduced $\chi^2$. Using resolution $\Delta\ell = 15$ yields even more modes with large $\chi^2$, because the cross spectra are able to extract features due to systematics which were not discernible with $\Delta\ell = 21$. In the next section, all power spectra calculated with extended mode projection were computed by projecting out modes with $\chi^2>1$. One can adjust this parameter to set the accepted level of correlation between with the data and the templates, above which modes are considered as contaminated by systematics and are projected out. One or multiple values for the $\chi^2$ cut should be decided before the analysis, so that the systematics mitigation relies on objective criteria, and is performed in a blind fashion. Alternatively, iterating and refining the $\chi^2$ cut during the analysis could improve the systematics mitigation, but may also yield fine-tuning and over-processing of the data if the results do not conform to expectation (confirmation bias), jeopardising the search for new physics. 

We finally note that previous analyses of SDSS quasars (\eg~\citealt{PullenHirata2012}) used samples with redshift distributions that were not overlapping: therefore their cross-power spectra could be used as null tests, \ie any systematics in common between the quasar samples would be detectable in the cross-spectra between redshift bins, which should be consistent with zero in the absence of systematics. In this work, the redshift distributions of the quasar samples overlap, some of them significantly, implying that the cross-spectra will contain cosmological information. We will demonstrate in the next section and the companion paper \citep{LeistedtPeiris:2014:XDQSO_png} that the fourth sample is not only the one with the most constraining power for PNG, but also the one with the redshift distribution overlapping the least with the other samples: its cross-power spectra can thus be used for approximate null tests.

\subsection{Power spectrum measurements}

\figref{fig:resultsnomp} shows the results of the QML estimator applied to the quasar samples in three settings: Masks 1 and 2 without mode projection, and Mask 2 with mode projection of the base set of systematics templates. The black lines show our fiducial theoretical predictions, detailed in the previous section. The shaded band indicates the zone spanned by the theory curves when varying PNG in $-50 < f_{\rm NL} <50$. Note that all spectra are dimensionless, since they relate to overdensity maps. At first examination, we see that all measured power spectra exhibit significant excess power on a range of scales compared to the theoretical expectation. In particular, the first and the fourth samples (the low and high redshift quasar samples) are in significant disagreement with the theoretical curve. The quasar cross-power spectra also exhibit power excesses, pointing to the presence of significant systematics and spatially-varying stellar contamination in these samples. Qualitatively, the discrepancy between the measurements and the prediction agrees with the levels of contamination measured by the null test $\chi^2$ histograms of the previous section, indicating that at least some of the systematics responsible for the excesses are included in the sets of templates prepared for mode projection.

Indeed, the third set of spectra of \figref{fig:resultsnomp} show that using the basic mode projection significantly decreases the power excesses in the auto-spectra, especially that of the fourth sample. The auto-power spectra are now subject to an offset compared to the theoretical predictions. Although this could be due to inadequate cosmological or observational parameters in the predictions, such as underestimated biases, significant excess power in the cross-power spectra indicates that part of the remaining discrepancies must be due to systematics. In the formalism of Section~\ref{sec:extmodeproj}, this also demonstrates that a linear model of the basic set of systematics ($\sim100$ uncorrelated modes) is insufficient to describe the contamination signal and spurious clustering. Moreover, since the error bars of \figref{fig:resultsnomp} only include cosmic variance and shot noise, we deduce that the contamination level is so large that it prevents any reliable measurement of the quasar bias and PNG. In fact, constraining\footnote{The error bars of \figref{fig:resultsnomp} use the fiducial priors and are included to illustrate the large discrepancies and power excesses due to systematics. To exploit these power spectra in a likelihood analysis, one would need to recompute the covariance matrices around the best-fit theoretical power spectra, which were done for the cleanest power spectrum estimates of \figref{fig:resultsemp}.} PNG from the measurements of \figref{fig:resultsnomp} (ignoring the null test cross-spectra and thus the evidence for high levels of systematics) would yield very large levels of PNG, significantly above the limits set by previous studies of SDSS quasars, luminous red galaxies (LRGs), and in the CMB (\eg \citealt{Giannantonio2013png, Planck2013nongaussianity}). The purpose of extended mode projection is precisely to address these issues, and reduce the levels of systematics below the statistical uncertainties by marginalising over the most contaminated modes. 

\figref{fig:resultsemp} shows three sets of estimates obtained when using extended mode projection with the more restrictive mask (Mask 2). The first set of spectra are calculated using band powers of width $\Delta\ell=21$ for both quasar spectra and null tests, the third set uses $\Delta\ell=15$ for both, and the second set is an intermediate setting where the quasar spectra use $\Delta\ell=15$ and the null tests $\Delta\ell=21$. This hybrid case is to check the robustness of the final spectra, since we observed in \figref{fig:chi2distribs} that the null tests with $\Delta\ell=21$ and $\Delta\ell=15$ yielded slightly different $\chi^2$ distributions for the first and fourth (most contaminated) quasar samples. 

Firstly, we see that the three settings are in good overall agreement with each other, and also with the fiducial theoretical predictions, as demonstrated by the $\chi^2$ values presented in Table~\ref{chi2vals_emp}. Indeed, despite a few band-powers that seem discrepant with the theory predictions at the $\sim 1-2\sigma$ level when examining \figref{fig:resultsemp}, the values of the probability to exceed (PTE) show that the fiducial spectra are good fits to the measurements. This indicates that extended mode projection succeeds at mitigating the most important systematics and reduce the  contamination to a level below the statistical uncertainties. Interestingly, the multipole resolution $\Delta\ell=21$ yields better $\chi^2$ than $\Delta\ell=15$. This can be explained by the fact that some systematics had their correlation signatures averaged and smoothed out at $\Delta\ell=21$, but resolved at $\Delta\ell=15$. In other words, greater multipole resolution generally uncovers more systematics, and may require the use of more numerous or more complex combinations of systematics templates. But for the three sets of spectra shown in \figref{fig:resultsemp}, none of the low-$\ell$ band powers in the auto- and cross-power spectra exhibit evidence for systematics. Therefore we conclude that they are suitable for cosmological inference and for constraining PNG, which produces a signal that is more pronounced on large scales. Indeed, the error bars and shaded bands of \figref{fig:resultsemp} demonstrate that the clean power spectra will be able to constrain PNG.

\begin{table}
	\centering
	\caption{The chi-square values for the power spectra measured using extended mode projection with Mask 2 with $\chi^2_{\rm cut} = 1.0$, presented in \figref{fig:resultsemp}. The theory curves use {\it Planck} $\Lambda$CDM cosmology, $\fnl=0$, and fiducial linear bias $b(z) = 1 + [(1+z)/2.5]^5$. The numbers of degrees of freedom are $\nu-p=7$ and $\nu-p=9$ for $\Delta\ell=21$ and $\Delta\ell=15$, respectively. The probability to exceed (PTE) the observed chi squares are shown in parentheses. Note that PTE $<1\%$ corresponds to $\chi^2_{7} = 18.4$ and $\chi^2_{9} = 21.6$. }
	\label{chi2vals_emp}
	\begin{tabular}{lccc}
		$\mCl$ estimator:	&	$\Delta\ell=21$	&	$\Delta\ell=15$	&	$\Delta\ell=15$	\\
		Ext. mp null tests: 	&	$\Delta\ell=21$	&	$\Delta\ell=21$	&	$\Delta\ell=15$	\\\hline
Bin 1  auto  & 4.81 (0.682)  & 7.42 (0.593)  & 7.66 (0.568) \\
Bin 2  $\times$ Bin 1  & 8.15 (0.319)  & 6.13 (0.727)  & 5.67 (0.772) \\
Bin 2  auto  & 3.89 (0.792)  & 4.79 (0.852)  & 6.23 (0.716) \\
Bin 3  $\times$ Bin 1  & 7.91 (0.341)  & 12.62 (0.180)  & 10.60 (0.304) \\
Bin 3  $\times$ Bin 2  & 2.19 (0.948)  & 4.02 (0.909)  & 9.50 (0.392) \\
Bin 3  auto  & 0.76 (0.997)  & 6.39 (0.699)  & 9.16 (0.423) \\
Bin 4  $\times$ Bin 1  & 8.43 (0.296)  & 13.60 (0.137)  & 10.71 (0.296) \\
Bin 4  $\times$ Bin 2  & 9.33 (0.230)  & 13.00 (0.162)  & 17.24 (0.045) \\
Bin 4  $\times$ Bin 3  & 5.83 (0.559)  & 7.55 (0.579)  & 12.94 (0.165) \\
Bin 4  auto  & 3.44 (0.8411)  & 3.14 (0.958)  & 4.07 (0.906) \\ \hline
	\end{tabular}
\end{table}


\section{Conclusion}

Photometric quasar surveys are deep and span extended redshift ranges, which allows us to probe the largest scales of the universe, and therefore test physics which is not well constrained by galaxy surveys. In particular, they can be used to constrain PNG, which is expected to enhance quasar clustering on large scales and leave a characteristic scale-dependent signature in the quasar bias. However, this requires accurate power spectrum measurements, which are compromised by the presence of numerous observational systematics, creating spurious correlations which can mimic the signatures of new physics.
 
We have introduced the {\it extended mode projection} technique to robustly mitigate the impact of large numbers of  systematics when estimating angular power spectra, and applied it to the photometric quasars from the SDSS XDQSOz catalogue. This technique only relies on the ability to map known and potential sources of systematics on the sky, and cross-correlate them with the data of interest. Previous studies of XDQSOz data required stringent quality and sky cuts, and even the removal of band-powers in order to avoid excessive contamination by systematics. In our analysis, we have used minimal sky cuts, and applied the extended mode projection approach using a large number of systematics templates. Mode projection is equivalent to a Bayesian marginalisation over the amplitudes of the modes of the contamination model when estimating the power spectra. The base set of templates included $220$ potential systematics found in the SDSS database, and we have also included products of pairs of templates, leading to a total of $\sim 22,000$ systematics templates, yielding a non-linear model for the contamination signal. We have then decorrelated the systematics, and cross-correlated the resulting $\sim 3,700$  orthogonal modes with the quasar samples to carry out null tests and detect the modes which most likely create spurious correlations in the data. We have finally estimated clean quasar power spectra by projecting out the modes which yielded reduced $\chi^2>1$ for the cross-spectrum null tests. Our pool of systematics and resulting contamination model was very general, and the sky masks minimal; thus, the reduced $\chi^2$ cut is the only tuneable parameter in the extended mode projection approach. Our approach is therefore based on the principles of blind analysis, since projecting out modes with reduced $\chi^2>1$ is a simple and pre-selected criterion for the accepted level of correlation between the systematics and the maps, which does not depend on the intrinsic clustering of quasars.  Using various settings for both the power spectrum estimation and the systematics mitigation, we have tested that the power spectrum measurements are robust to these choices, and consistent with the theoretical predictions. In a companion paper \citep{LeistedtPeiris:2014:XDQSO_png}, we show that these spectra, used in a combined likelihood function, yield stringent and robust constraints on PNG and on the bias of quasars. In particular, the constraints separately derived using the auto- and cross-spectra  are consistent with each other, and robust to the underlying model and assumptions, for example to the uncertainties in the redshift distributions of the samples. This demonstrates that the remaining levels of correlations created by systematics are below the statistical uncertainties, and that the quasar power spectra are suitable for use in cosmological inferences. 

Future galaxy and quasar survey data will reach unprecedented precision, and will require accurate mitigation of large numbers of systematics. For instance DES, {\it Euclid} \citep{Amendola:2012ys}, and LSST \citep{Abell:2009aa} will observe hundreds of millions of objects, and probe extended redshift ranges and finer angular scales. This makes them very promising for testing new physics beyond the standard cosmological model, such as PNG, the neutrino sector, dark energy phenomenology, and modifications to General Relativity. At such precision levels, and given that these new physics signatures are typically small deviations from the standard model, power spectrum measurements will be highly sensitive to any systematics. The extended mode projection framework is a good candidate for mitigating such systematics in a robust and blind fashion, while extracting as much information as possible from the hard-won data.

\section{Acknowledgements}

We are grateful to Jo Bovy, David Hogg and Adam Myers for sharing the XDQSOz catalogue with us. We also thank Daniel Mortlock,  Nina Roth, Filipe Abdalla, Aur\'elien~Benoit-L\'evy, and Blake Sherwin,  for useful discussions and comments. BL is supported by the Perren Fund and the IMPACT Fund. HVP is supported by STFC and the European Research Council under the European Community's Seventh Framework Programme (FP7/2007- 2013) / ERC grant agreement no 306478-CosmicDawn. 

We acknowledge use of the following public software packages: \textsc{healpix} \citep{healpix1} and \textsc{camb\_sources} \citep{challinorlewis2011cambsources}. We acknowledge use of the Legacy Archive for Microwave Background Data Analysis (LAMBDA). Support for LAMBDA is provided by the NASA Office of Space Science. 

This work is based on observations obtained with SDSS. Funding for the SDSS and SDSS-II has been provided by the Alfred P. Sloan Foundation, the Participating Institutions, the National Science Foundation, the U.S. Department of Energy, the National Aeronautics and Space Administration, the Japanese Monbukagakusho, the Max Planck Society, and the Higher Education Funding Council for England. The SDSS Web Site is http://www.sdss.org/. The SDSS is managed by the Astrophysical Research Consortium for the Participating Institutions. The Participating Institutions are the American Museum of Natural History, Astrophysical Institute Potsdam, University of Basel, University of Cambridge, Case Western Reserve University, University of Chicago, Drexel University, Fermilab, the Institute for Advanced Study, the Japan Participation Group, Johns Hopkins University, the Joint Institute for Nuclear Astrophysics, the Kavli Institute for Particle Astrophysics and Cosmology, the Korean Scientist Group, the Chinese Academy of Sciences (LAMOST), Los Alamos National Laboratory, the Max-Planck-Institute for Astronomy (MPIA), the Max-Planck-Institute for Astrophysics (MPA), New Mexico State University, Ohio State University, University of Pittsburgh, University of Portsmouth, Princeton University, the United States Naval Observatory, and the University of Washington.

\footnotesize{
  \bibliographystyle{mn2e}
\providecommand{\eprint}[1]{\href{http://arxiv.org/abs/#1}{arXiv:#1}}	
  \bibliography{bib}

\begin{thebibliography}{68}
\expandafter\ifx\csname natexlab\endcsname\relax\def\natexlab#1{#1}\fi

\bibitem[{Abell {et~al}\mbox{.}(2009)Abell {et~al.}}]{Abell:2009aa}
Abell P.~A., {et~al.}, 2009, ArXiv e-prints

\bibitem[{{Agarwal} {et~al}\mbox{.}(2013){Agarwal}, {Ho}, {Myers}, {Seo},
  {Ross}, {Bahcall}, {Brinkmann}, {Eisenstein}, {Muna},
  {Palanque-Delabrouille}, {P{\^a}ris}, {Petitjean}, {Schneider},
  {Streblyanska}, {Weaver}, \& {Y{\`e}che}}]{agarwalho2013sys}
{Agarwal} N. {et~al.}, 2013, ArXiv e-prints

\bibitem[{{Agarwal}, {Ho} \& {Shandera}(2013){Agarwal}, {Ho}, \&
  {Shandera}}]{agarwalho2013xdqsoz}
{Agarwal} N., {Ho} S., {Shandera} S., 2013, ArXiv e-prints

\bibitem[{Amendola {et~al}\mbox{.}(2013)Amendola {et~al.}}]{Amendola:2012ys}
Amendola L., {et~al.}, 2013, Living Rev.Rel., 16, 6

\bibitem[{{Bovy} {et~al}\mbox{.}(2011){Bovy}, {Hennawi}, {Hogg}, {Myers},
  {Kirkpatrick}, {Schlegel}, {Ross}, {Sheldon}, {McGreer}, {Schneider}, \&
  {Weaver}}]{Bovy2010xdqso}
{Bovy} J. {et~al.}, 2011, Astrophys.\ J., 729, 141

\bibitem[{{Bovy}, {Hogg} \& {Roweis}(2011){Bovy}, {Hogg}, \&
  {Roweis}}]{Bovy2011xd}
{Bovy} J., {Hogg} D.~W., {Roweis} S.~T., 2011, Annals of Applied Statistics, 5,
  1657

\bibitem[{{Bovy} {et~al}\mbox{.}(2012){Bovy}, {Myers}, {Hennawi}, {Hogg},
  {McMahon}, {Schiminovich}, {Sheldon}, {Brinkmann}, {Schneider}, \&
  {Weaver}}]{bovy2012xdqsoz}
{Bovy} J. {et~al.}, 2012, Astrophys.\ J., 749, 41

\bibitem[{{Challinor} \& {Lewis}(2011)}]{challinorlewis2011cambsources}
{Challinor} A., {Lewis} A., 2011, Phys.\ Rev.\ D., 84, 043516

\bibitem[{{Dalal} {et~al}\mbox{.}(2008){Dalal}, {Dor{\'e}}, {Huterer}, \&
  {Shirokov}}]{Dalal2008png}
{Dalal} N., {Dor{\'e}} O., {Huterer} D., {Shirokov} A., 2008, Phys.\ Rev.\ D.,
  77, 123514

\bibitem[{{Dawson} {et~al}\mbox{.}(2013){Dawson}, {Schlegel}, {Ahn},
  {Anderson}, {Aubourg}, {Bailey}, {Barkhouser}, {Bautista}, {Beifiori},
  {Berlind}, {Bhardwaj}, {Bizyaev}, {Blake}, {Blanton}, {Blomqvist}, {Bolton},
  {Borde}, {Bovy}, {Brandt}, {Brewington}, {Brinkmann}, {Brown}, {Brownstein},
  {Bundy}, {Busca}, {Carithers}, {Carnero}, {Carr}, {Chen}, {Comparat},
  {Connolly}, {Cope}, {Croft}, {Cuesta}, {da Costa}, {Davenport}, {Delubac},
  {de Putter}, {Dhital}, {Ealet}, {Ebelke}, {Eisenstein}, {Escoffier}, {Fan},
  {Filiz Ak}, {Finley}, {Font-Ribera}, {G{\'e}nova-Santos}, {Gunn}, {Guo},
  {Haggard}, {Hall}, {Hamilton}, {Harris}, {Harris}, {Ho}, {Hogg}, {Holder},
  {Honscheid}, {Huehnerhoff}, {Jordan}, {Jordan}, {Kauffmann}, {Kazin},
  {Kirkby}, {Klaene}, {Kneib}, {Le Goff}, {Lee}, {Long}, {Loomis}, {Lundgren},
  {Lupton}, {Maia}, {Makler}, {Malanushenko}, {Malanushenko}, {Mandelbaum},
  {Manera}, {Maraston}, {Margala}, {Masters}, {McBride}, {McDonald}, {McGreer},
  {McMahon}, {Mena}, {Miralda-Escud{\'e}}, {Montero-Dorta}, {Montesano},
  {Muna}, {Myers}, {Naugle}, {Nichol}, {Noterdaeme}, {Nuza}, {Olmstead},
  {Oravetz}, {Oravetz}, {Owen}, {Padmanabhan}, {Palanque-Delabrouille}, {Pan},
  {Parejko}, {P{\^a}ris}, {Percival}, {P{\'e}rez-Fournon},
  {P{\'e}rez-R{\`a}fols}, {Petitjean}, {Pfaffenberger}, {Pforr}, {Pieri},
  {Prada}, {Price-Whelan}, {Raddick}, {Rebolo}, {Rich}, {Richards}, {Rockosi},
  {Roe}, {Ross}, {Ross}, {Rossi}, {Rubi{\~n}o-Martin}, {Samushia},
  {S{\'a}nchez}, {Sayres}, {Schmidt}, {Schneider}, {Sc{\'o}ccola}, {Seo},
  {Shelden}, {Sheldon}, {Shen}, {Shu}, {Slosar}, {Smee}, {Snedden}, {Stauffer},
  {Steele}, {Strauss}, {Streblyanska}, {Suzuki}, {Swanson}, {Tal}, {Tanaka},
  {Thomas}, {Tinker}, {Tojeiro}, {Tremonti}, {Vargas Maga{\~n}a}, {Verde},
  {Viel}, {Wake}, {Watson}, {Weaver}, {Weinberg}, {Weiner}, {West}, {White},
  {Wood-Vasey}, {Yeche}, {Zehavi}, {Zhao}, \& {Zheng}}]{dawson2013boss}
{Dawson} K.~S. {et~al.}, 2013, Astrophys.\ J., 145, 10

\bibitem[{{Efstathiou}(2004{\natexlab{a}})}]{Efsta2003}
{Efstathiou} G., 2004{\natexlab{a}}, Mon.\ Not.\ Roy.\ Astron.\ Soc., 348, 885

\bibitem[{{Efstathiou}(2004{\natexlab{b}})}]{Efsta2004}
{Efstathiou} G., 2004{\natexlab{b}}, Mon.\ Not.\ Roy.\ Astron.\ Soc., 349, 603

\bibitem[{{Efstathiou}(2006)}]{Efsta2006}
{Efstathiou} G., 2006, Mon.\ Not.\ Roy.\ Astron.\ Soc., 370, 343

\bibitem[{Fan {et~al}\mbox{.}(2006)Fan, Strauss, Becker, White, Gunn,
  {et~al.}}]{Fan:2005es}
Fan X.-H., Strauss M.~A., Becker R.~H., White R.~L., Gunn J.~E., {et~al.},
  2006, Astron.J., 132, 117

\bibitem[{{Feeney}, {Peiris} \& {Pontzen}(2011){Feeney}, {Peiris}, \&
  {Pontzen}}]{FPP11}
{Feeney} S.~M., {Peiris} H.~V., {Pontzen} A., 2011, Phys.\ Rev.\ D., 84, 103002

\bibitem[{Giannantonio {et~al}\mbox{.}(2012)Giannantonio, Crittenden, Nichol,
  \& Ross}]{Giannantonio01112012}
Giannantonio T., Crittenden R., Nichol R., Ross A.~J., 2012, Mon.\ Not.\ Roy.\
  Astron.\ Soc., 426, 2581

\bibitem[{{Giannantonio} \& {Percival}(2013)}]{Giannantonio2013crosscmblss}
{Giannantonio} T., {Percival} W.~J., 2013, ArXiv e-prints

\bibitem[{{Giannantonio} {et~al}\mbox{.}(2013){Giannantonio}, {Ross},
  {Percival}, {Crittenden}, {Bacher}, {Kilbinger}, {Nichol}, \&
  {Weller}}]{Giannantonio2013png}
{Giannantonio} T., {Ross} A.~J., {Percival} W.~J., {Crittenden} R., {Bacher}
  D., {Kilbinger} M., {Nichol} R., {Weller} J., 2013, ArXiv e-prints

\bibitem[{{G{\'o}rski} {et~al}\mbox{.}(2005){G{\'o}rski}, {Hivon}, {Banday},
  {Wandelt}, {Hansen}, {Reinecke}, \& {Bartelmann}}]{healpix1}
{G{\'o}rski} K.~M., {Hivon} E., {Banday} A.~J., {Wandelt} B.~D., {Hansen}
  F.~K., {Reinecke} M., {Bartelmann} M., 2005, Astrophys.\ J., 622, 759

\bibitem[{{Gruetjen} \& {Shellard}(2012)}]{GruetjenAndShellard2012}
{Gruetjen} H.~F., {Shellard} E.~P.~S., 2012, ArXiv e-prints

\bibitem[{{Gunn} {et~al}\mbox{.}(2006){Gunn}, {Siegmund}, {Mannery}, {Owen},
  {Hull}, {Leger}, {Carey}, {Knapp}, {York}, {Boroski}, {Kent}, {Lupton},
  {Rockosi}, {Evans}, {Waddell}, {Anderson}, {Annis}, {Barentine}, {Bartoszek},
  {Bastian}, {Bracker}, {Brewington}, {Briegel}, {Brinkmann}, {Brown}, {Carr},
  {Czarapata}, {Drennan}, {Dombeck}, {Federwitz}, {Gillespie}, {Gonzales},
  {Hansen}, {Harvanek}, {Hayes}, {Jordan}, {Kinney}, {Klaene}, {Kleinman},
  {Kron}, {Kresinski}, {Lee}, {Limmongkol}, {Lindenmeyer}, {Long}, {Loomis},
  {McGehee}, {Mantsch}, {Neilsen}, {Neswold}, {Newman}, {Nitta}, {Peoples},
  {Pier}, {Prieto}, {Prosapio}, {Rivetta}, {Schneider}, {Snedden}, \&
  {Wang}}]{Gunn2006}
{Gunn} J.~E. {et~al.}, 2006, Astrophys.\ J., 131, 2332

\bibitem[{{Hamilton} \& {Tegmark}(2004)}]{Hamilton2004mangle}
{Hamilton} A.~J.~S., {Tegmark} M., 2004, Mon.\ Not.\ Roy.\ Astron.\ Soc., 349,
  115

\bibitem[{{Ho} {et~al}\mbox{.}(2013){Ho}, {Agarwal}, {Myers}, {Lyons},
  {Disbrow}, {Seo}, {Ross}, {Hirata}, {Padmanabhan}, {O'Connell}, {Huff},
  {Schlegel}, {Slosar}, {Weinberg}, {Strauss}, {Ross}, {Schneider}, {Bahcall},
  {Brinkmann}, {Palanque-Delabrouille}, \& {Y{\`e}che}}]{hoagarwal2013xdqsoz}
{Ho} S. {et~al.}, 2013, ArXiv e-prints

\bibitem[{{Ho} {et~al}\mbox{.}(2012){Ho}, {Cuesta}, {Seo}, {de Putter}, {Ross},
  {White}, {Padmanabhan}, {Saito}, {Schlegel}, {Schlafly}, {Seljak},
  {Hern{\'a}ndez-Monteagudo}, {S{\'a}nchez}, {Percival}, {Blanton}, {Skibba},
  {Schneider}, {Reid}, {Mena}, {Viel}, {Eisenstein}, {Prada}, {Weaver},
  {Bahcall}, {Bizyaev}, {Brewinton}, {Brinkman}, {Nicolaci da Costa}, {Gott},
  {Malanushenko}, {Malanushenko}, {Nichol}, {Oravetz}, {Pan},
  {Palanque-Delabrouille}, {Ross}, {Simmons}, {de Simoni}, {Snedden}, \&
  {Yeche}}]{ho2012cosmoweights}
{Ho} S. {et~al.}, 2012, Astrophys.\ J., 761, 14

\bibitem[{{Huterer}, {Cunha} \& {Fang}(2013){Huterer}, {Cunha}, \&
  {Fang}}]{Huterer2012calibrationerrors}
{Huterer} D., {Cunha} C.~E., {Fang} W., 2013, Mon.\ Not.\ Roy.\ Astron.\ Soc.

\bibitem[{{Karagiannis}, {Shanks} \& {Ross}(2013){Karagiannis}, {Shanks}, \&
  {Ross}}]{Karagiannis2013}
{Karagiannis} D., {Shanks} T., {Ross} N.~P., 2013, ArXiv e-prints

\bibitem[{{Knox}, {Bond} \& {Jaffe}(1998){Knox}, {Bond}, \& {Jaffe}}]{KBJ98}
{Knox} L., {Bond} J.~R., {Jaffe} A.~H., 1998, in Eighteenth Texas Symposium on
  Relativistic Astrophysics, {A.~V.~Olinto, J.~A.~Frieman, \& D.~N.~Schramm},
  ed., p. 282

\bibitem[{{Komatsu} {et~al}\mbox{.}(2009){Komatsu}, {Afshordi}, {Bartolo},
  {Baumann}, {Bond}, {Buchbinder}, {Byrnes}, {Chen}, {Chung}, {Cooray},
  {Creminelli}, {Dalal}, {Dore}, {Easther}, {Frolov}, {Khoury}, {Kinney},
  {Kofman}, {Koyama}, {Leblond}, {Lehners}, {Lidsey}, {Liguori}, {Lim},
  {Linde}, {Lyth}, {Maldacena}, {Matarrese}, {McAllister}, {McDonald},
  {Mukohyama}, {Ovrut}, {Peiris}, {Riotto}, {Rodrigues}, {Sasaki},
  {Scoccimarro}, {Seery}, {Sefusatti}, {Smith}, {Starobinsky}, {Steinhardt},
  {Takahashi}, {Tegmark}, {Tolley}, {Verde}, {Wandelt}, {Wands}, {Weinberg},
  {Wyman}, {Yadav}, \& {Zaldarriaga}}]{2009astro2010S158K}
{Komatsu} E. {et~al.}, 2009, in Astronomy, Vol. 2010, astro2010: The Astronomy
  and Astrophysics Decadal Survey, p. 158

\bibitem[{{Leistedt} \& {Peiris}(2014)}]{LeistedtPeiris:2014:XDQSO_png}
{Leistedt} B., {Peiris} H.~V., 2014, ArXiv e-prints

\bibitem[{{Leistedt} {et~al}\mbox{.}(2013){Leistedt}, {Peiris}, {Mortlock},
  {Benoit-L{\'e}vy}, \& {Pontzen}}]{Leistedt2013excessdr6}
{Leistedt} B., {Peiris} H.~V., {Mortlock} D.~J., {Benoit-L{\'e}vy} A.,
  {Pontzen} A., 2013, Mon.\ Not.\ Roy.\ Astron.\ Soc., 435, 1857

\bibitem[{{LoVerde} {et~al}\mbox{.}(2008){LoVerde}, {Miller}, {Shandera}, \&
  {Verde}}]{loverde2008}
{LoVerde} M., {Miller} A., {Shandera} S., {Verde} L., 2008, Astrophys.\ J.\
  Lett., 4, 14

\bibitem[{{Matarrese} \& {Verde}(2008)}]{matarrese2008}
{Matarrese} S., {Verde} L., 2008, Astrophys.\ J.\ Lett., 677, L77

\bibitem[{Matthews \& Newman(2010)}]{Matthews:2010an}
Matthews D.~J., Newman J.~A., 2010, Astrophys.J., 721, 456

\bibitem[{McQuinn \& White(2013)}]{McQuinn:2013ib}
McQuinn M., White M., 2013, Mon.Not.Roy.Astron.Soc., 433, 2857

\bibitem[{{Myers} {et~al}\mbox{.}(2007{\natexlab{a}}){Myers}, {Brunner},
  {Nichol}, {Richards}, {Schneider}, \& {Bahcall}}]{Myers2007one}
{Myers} A.~D., {Brunner} R.~J., {Nichol} R.~C., {Richards} G.~T., {Schneider}
  D.~P., {Bahcall} N.~A., 2007{\natexlab{a}}, Astrophys.\ J., 658, 85

\bibitem[{{Myers} {et~al}\mbox{.}(2007{\natexlab{b}}){Myers}, {Brunner},
  {Richards}, {Nichol}, {Schneider}, \& {Bahcall}}]{Myers2007two}
{Myers} A.~D., {Brunner} R.~J., {Richards} G.~T., {Nichol} R.~C., {Schneider}
  D.~P., {Bahcall} N.~A., 2007{\natexlab{b}}, Astrophys.\ J., 658, 99

\bibitem[{{Myers} {et~al}\mbox{.}(2006){Myers}, {Brunner}, {Richards},
  {Nichol}, {Schneider}, {Vanden Berk}, {Scranton}, {Gray}, \&
  {Brinkmann}}]{Myers2006first}
{Myers} A.~D. {et~al.}, 2006, Astrophys.\ J., 638, 622

\bibitem[{{Outram} {et~al}\mbox{.}(2003){Outram}, {Hoyle}, {Shanks}, {Croom},
  {Boyle}, {Miller}, {Smith}, \& {Myers}}]{Outram20032dfqso}
{Outram} P.~J., {Hoyle} F., {Shanks} T., {Croom} S.~M., {Boyle} B.~J., {Miller}
  L., {Smith} R.~J., {Myers} A.~D., 2003, Mon.\ Not.\ Roy.\ Astron.\ Soc., 342,
  483

\bibitem[{{P{\^a}ris} {et~al}\mbox{.}(2012){P{\^a}ris}, {Petitjean}, {Aubourg},
  {Bailey}, {Ross}, {Myers}, {Strauss}, {Anderson}, {Arnau}, {Bautista},
  {Bizyaev}, {Bolton}, {Bovy}, {Brandt}, {Brewington}, {Browstein}, {Busca},
  {Capellupo}, {Carithers}, {Croft}, {Dawson}, {Delubac}, {Ebelke},
  {Eisenstein}, {Engelke}, {Fan}, {Filiz Ak}, {Finley}, {Font-Ribera}, {Ge},
  {Gibson}, {Hall}, {Hamann}, {Hennawi}, {Ho}, {Hogg}, {Ivezi{\'c}}, {Jiang},
  {Kimball}, {Kirkby}, {Kirkpatrick}, {Lee}, {Le Goff}, {Lundgren}, {MacLeod},
  {Malanushenko}, {Malanushenko}, {Maraston}, {McGreer}, {McMahon},
  {Miralda-Escud{\'e}}, {Muna}, {Noterdaeme}, {Oravetz},
  {Palanque-Delabrouille}, {Pan}, {Perez-Fournon}, {Pieri}, {Richards},
  {Rollinde}, {Sheldon}, {Schlegel}, {Schneider}, {Slosar}, {Shelden}, {Shen},
  {Simmons}, {Snedden}, {Suzuki}, {Tinker}, {Viel}, {Weaver}, {Weinberg},
  {White}, {Wood-Vasey}, \& {Y{\`e}che}}]{paris2012bossqsodr9}
{P{\^a}ris} I. {et~al.}, 2012, Astron.\ \& Astrophys., 548, A66

\bibitem[{{Planck
  Collaboration}(2013{\natexlab{a}})}]{Planck2013cosmologicalparams}
{Planck Collaboration}, 2013{\natexlab{a}}, ArXiv e-prints

\bibitem[{{Planck
  Collaboration}(2013{\natexlab{b}})}]{Planck2013nongaussianity}
{Planck Collaboration}, 2013{\natexlab{b}}, ArXiv e-prints

\bibitem[{{Planck Collaboration} {et~al}\mbox{.}(2013){Planck Collaboration},
  {Abergel}, {Ade}, {Aghanim}, {Alina}, {Alves}, {Aniano}, {Arnaud}, {Ashdown},
  {Aumont}, {Baccigalupi}, {Banday}, {Barreiro}, {Bartlett}, {Battaner},
  {Benabed}, {Benoit-L{\'e}vy}, {Bernard}, {Bersanelli}, {Bielewicz}, {Bobin},
  {Bonaldi}, {Bond}, {Bouchet}, {Boulanger}, {Burigana}, {Cardoso}, {Catalano},
  {Chamballu}, {Chiang}, {Christensen}, {Clements}, {Colombi}, {Colombo},
  {Couchot}, {Crill}, {Cuttaia}, {Danese}, {Davies}, {Davis}, {de Bernardis},
  {de Rosa}, {de Zotti}, {Delabrouille}, {D{\'e}sert}, {Dickinson}, {Diego},
  {Dole}, {Donzelli}, {Dor{\'e}}, {Douspis}, {Dupac}, {Efstathiou},
  {En{\ss}lin}, {Eriksen}, {Falgarone}, {Finelli}, {Forni}, {Frailis},
  {Franceschi}, {Galeotta}, {Ganga}, {Ghosh}, {Giard}, {Giraud-H{\'e}raud},
  {Gonz{\'a}lez-Nuevo}, {G{\'o}rski}, {Gregorio}, {Gruppuso}, {Guillet},
  {Hansen}, {Harrison}, {Helou}, {Henrot-Versill{\'e}},
  {Hern{\'a}ndez-Monteagudo}, {Herranz}, {Hildebrandt}, {Hivon}, {Hobson},
  {Holmes}, {Hornstrup}, {Hovest}, {Huffenberger}, {Jaffe}, {Jaffe}, {Joncas},
  {Jones}, {Jones}, {Juvela}, {Kalberla}, {Keih{\"a}nen}, {Kerp}, {Keskitalo},
  {Kisner}, {Kneissl}, {Knoche}, {Kunz}, {Kurki-Suonio}, {Lagache},
  {L{\"a}hteenm{\"a}ki}, {Lamarre}, {Lasenby}, {Leonardi}, {Levrier},
  {Liguori}, {Lilje}, {Linden-V{\o}rnle}, {L{\'o}pez-Caniego}, {Lubin},
  {Mac{\'{\i}}as-P{\'e}rez}, {Maffei}, {Maino}, {Mandolesi}, {Maris},
  {Marshall}, {Martin}, {Mart{\'{\i}}nez-Gonz{\'a}lez}, {Masi}, {Massardi},
  {Matarrese}, {Mazzotta}, {Melchiorri}, {Mendes}, {Mennella}, {Migliaccio},
  {Mitra}, {Miville-Desch{\^e}nes}, {Moneti}, {Montier}, {Morgante},
  {Mortlock}, {Munshi}, {Murphy}, {Naselsky}, {Nati}, {Natoli}, {Noviello},
  {Novikov}, {Novikov}, {Oxborrow}, {Pagano}, {Pajot}, {Paoletti}, {Pasian},
  {Perdereau}, {Perotto}, {Perrotta}, {Piacentini}, {Piat}, {Pierpaoli},
  {Pietrobon}, {Plaszczynski}, {Pointecouteau}, {Polenta}, {Ponthieu}, {Popa},
  {Pratt}, {Prunet}, {Puget}, {Rachen}, {Reach}, {Rebolo}, {Reinecke},
  {Remazeilles}, {Renault}, {Ricciardi}, {Riller}, {Ristorcelli}, {Rocha},
  {Rosset}, {Roudier}, {Rusholme}, {Sandri}, {Savini}, {Spencer}, {Starck},
  {Sureau}, {Sutton}, {Suur-Uski}, {Sygnet}, {Tauber}, {Terenzi}, {Toffolatti},
  {Tomasi}, {Tristram}, {Tucci}, {Umana}, {Valenziano}, {Valiviita}, {Van
  Tent}, {Verstraete}, {Vielva}, {Villa}, {Wade}, {Wandelt}, {Winkel}, {Yvon},
  {Zacchei}, \& {Zonca}}]{Planck2003dust}
{Planck Collaboration} {et~al.}, 2013, ArXiv e-prints

\bibitem[{{Pontzen} \& {Peiris}(2010)}]{PP10}
{Pontzen} A., {Peiris} H.~V., 2010, Phys.\ Rev.\ D., 81, 103008

\bibitem[{{Porciani} \& {Norberg}(2006)}]{porcianiNorberg2006}
{Porciani} C., {Norberg} P., 2006, Mon.\ Not.\ Roy.\ Astron.\ Soc., 371, 1824

\bibitem[{{Pullen} \& {Hirata}(2012)}]{PullenHirata2012}
{Pullen} A.~R., {Hirata} C.~M., 2012, ArXiv e-prints

\bibitem[{{Richards} {et~al}\mbox{.}(2009){Richards}, {Myers}, {Gray},
  {Riegel}, {Nichol}, {Brunner}, {Szalay}, {Schneider}, \&
  {Anderson}}]{Richards2008rqcat}
{Richards} G.~T. {et~al.}, 2009, Astrophys.\ J.\ Supp., 180, 67

\bibitem[{{Richards} {et~al}\mbox{.}(2006){Richards}, {Strauss}, {Fan}, {Hall},
  {Jester}, {Schneider}, {Vanden Berk}, {Stoughton}, {Anderson}, {Brunner},
  {Gray}, {Gunn}, {Ivezi{\'c}}, {Kirkland}, {Knapp}, {Loveday}, {Meiksin},
  {Pope}, {Szalay}, {Thakar}, {Yanny}, {York}, {Barentine}, {Brewington},
  {Brinkmann}, {Fukugita}, {Harvanek}, {Kent}, {Kleinman}, {Krzesi{\'n}ski},
  {Long}, {Lupton}, {Nash}, {Neilsen}, {Nitta}, {Schlegel}, \&
  {Snedden}}]{Richards2006qlf}
{Richards} G.~T. {et~al.}, 2006, Astrophys.\ J., 131, 2766

\bibitem[{{Ross} {et~al}\mbox{.}(2011){Ross}, {Ho}, {Cuesta}, {Tojeiro},
  {Percival}, {Wake}, {Masters}, {Nichol}, {Myers}, {de Simoni}, {Seo},
  {Hern{\'a}ndez-Monteagudo}, {Crittenden}, {Blanton}, {Brinkmann}, {da Costa},
  {Guo}, {Kazin}, {Maia}, {Maraston}, {Padmanabhan}, {Prada}, {Ramos},
  {Sanchez}, {Schlafly}, {Schlegel}, {Schneider}, {Skibba}, {Thomas}, {Weaver},
  {White}, \& {Zehavi}}]{ross2011weights}
{Ross} A.~J. {et~al.}, 2011, Mon.\ Not.\ Roy.\ Astron.\ Soc., 417, 1350

\bibitem[{{Ross} {et~al}\mbox{.}(2013){Ross}, {Percival}, {Carnero}, {Zhao},
  {Manera}, {Raccanelli}, {Aubourg}, {Bizyaev}, {Brewington}, {Brinkmann},
  {Brownstein}, {Cuesta}, {da Costa}, {Eisenstein}, {Ebelke}, {Guo},
  {Hamilton}, {Maga{\~n}a}, {Malanushenko}, {Malanushenko}, {Maraston},
  {Montesano}, {Nichol}, {Oravetz}, {Pan}, {Prada}, {S{\'a}nchez}, {Samushia},
  {Schlegel}, {Schneider}, {Seo}, {Sheldon}, {Simmons}, {Snedden}, {Swanson},
  {Thomas}, {Tinker}, {Tojeiro}, \& {Zehavi}}]{ross2012png}
{Ross} A.~J. {et~al.}, 2013, Mon.\ Not.\ Roy.\ Astron.\ Soc., 428, 1116

\bibitem[{{Ross} {et~al}\mbox{.}(2012{\natexlab{a}}){Ross}, {Percival},
  {S{\'a}nchez}, {Samushia}, {Ho}, {Kazin}, {Manera}, {Reid}, {White},
  {Tojeiro}, {McBride}, {Xu}, {Wake}, {Strauss}, {Montesano}, {Swanson},
  {Bailey}, {Bolton}, {Dorta}, {Eisenstein}, {Guo}, {Hamilton}, {Nichol},
  {Padmanabhan}, {Prada}, {Schlegel}, {Maga{\~n}a}, {Zehavi}, {Blanton},
  {Bizyaev}, {Brewington}, {Cuesta}, {Malanushenko}, {Malanushenko}, {Oravetz},
  {Parejko}, {Pan}, {Schneider}, {Shelden}, {Simmons}, {Snedden}, \&
  {Zhao}}]{ross2012systematics}
{Ross} A.~J. {et~al.}, 2012{\natexlab{a}}, Mon.\ Not.\ Roy.\ Astron.\ Soc.,
  424, 564

\bibitem[{{Ross} {et~al}\mbox{.}(2012{\natexlab{b}}){Ross}, {Myers}, {Sheldon},
  {Y{\`e}che}, {Strauss}, {Bovy}, {Kirkpatrick}, {Richards}, {Aubourg},
  {Blanton}, {Brandt}, {Carithers}, {Croft}, {da Silva}, {Dawson},
  {Eisenstein}, {Hennawi}, {Ho}, {Hogg}, {Lee}, {Lundgren}, {McMahon},
  {Miralda-Escud{\'e}}, {Palanque-Delabrouille}, {P{\^a}ris}, {Petitjean},
  {Pieri}, {Rich}, {Roe}, {Schiminovich}, {Schlegel}, {Schneider}, {Slosar},
  {Suzuki}, {Tinker}, {Weinberg}, {Weyant}, {White}, \&
  {Wood-Vasey}}]{Ross2012dr9qsotarget}
{Ross} N.~P. {et~al.}, 2012{\natexlab{b}}, Astrophys.\ J.\ Supp., 199, 3

\bibitem[{{Ross} {et~al}\mbox{.}(2009){Ross}, {Shen}, {Strauss}, {Vanden Berk},
  {Connolly}, {Richards}, {Schneider}, {Weinberg}, {Hall}, {Bahcall}, \&
  {Brunner}}]{Ross2009specqsodr5}
{Ross} N.~P. {et~al.}, 2009, Astrophys.\ J., 697, 1634

\bibitem[{{Schlegel}, {Finkbeiner} \& {Davis}(1998){Schlegel}, {Finkbeiner}, \&
  {Davis}}]{Schlegel1998dust}
{Schlegel} D.~J., {Finkbeiner} D.~P., {Davis} M., 1998, Astrophys.\ J., 500,
  525

\bibitem[{{Schneider} {et~al}\mbox{.}(2010){Schneider}, {Richards}, {Hall},
  {Strauss}, {Anderson}, {Boroson}, {Ross}, {Shen}, {Brandt}, {Fan}, {Inada},
  {Jester}, {Knapp}, {Krawczyk}, {Thakar}, {Vanden Berk}, {Voges}, {Yanny},
  {York}, {Bahcall}, {Bizyaev}, {Blanton}, {Brewington}, {Brinkmann},
  {Eisenstein}, {Frieman}, {Fukugita}, {Gray}, {Gunn}, {Hibon}, {Ivezi{\'c}},
  {Kent}, {Kron}, {Lee}, {Lupton}, {Malanushenko}, {Malanushenko}, {Oravetz},
  {Pan}, {Pier}, {Price}, {Saxe}, {Schlegel}, {Simmons}, {Snedden}, {SubbaRao},
  {Szalay}, \& {Weinberg}}]{Schneider2010qsodr7cat}
{Schneider} D.~P. {et~al.}, 2010, Astrophys.\ J., 139, 2360

\bibitem[{{Serber} {et~al}\mbox{.}(2006){Serber}, {Bahcall}, {M{\'e}nard}, \&
  {Richards}}]{Serber2006}
{Serber} W., {Bahcall} N., {M{\'e}nard} B., {Richards} G., 2006, Astrophys.\
  J., 643, 68

\bibitem[{{Shen} {et~al}\mbox{.}(2007){Shen}, {Strauss}, {Oguri}, {Hennawi},
  {Fan}, {Richards}, {Hall}, {Gunn}, {Schneider}, {Szalay}, {Thakar}, {Vanden
  Berk}, {Anderson}, {Bahcall}, {Connolly}, \& {Knapp}}]{Shen2007specqso}
{Shen} Y. {et~al.}, 2007, Astrophys.\ J., 133, 2222

\bibitem[{{Sherwin} {et~al}\mbox{.}(2012){Sherwin}, {Das}, {Hajian}, {Addison},
  {Bond}, {Crichton}, {Devlin}, {Dunkley}, {Gralla}, {Halpern}, {Hill},
  {Hincks}, {Hughes}, {Huffenberger}, {Hlozek}, {Kosowsky}, {Louis},
  {Marriage}, {Marsden}, {Menanteau}, {Moodley}, {Niemack}, {Page}, {Reese},
  {Sehgal}, {Sievers}, {Sif{\'o}n}, {Spergel}, {Staggs}, {Switzer}, \&
  {Wollack}}]{Sherwin2012qsolensing}
{Sherwin} B.~D. {et~al.}, 2012, Phys.\ Rev.\ D., 86, 083006

\bibitem[{{Sievers} {et~al}\mbox{.}(2013){Sievers}, {Hlozek}, {Nolta},
  {Acquaviva}, {Addison}, {Ade}, {Aguirre}, {Amiri}, {Appel}, {Barrientos},
  {Battistelli}, {Battaglia}, {Bond}, {Brown}, {Burger}, {Calabrese},
  {Chervenak}, {Crichton}, {Das}, {Devlin}, {Dicker}, {Bertrand Doriese},
  {Dunkley}, {D{\"u}nner}, {Essinger-Hileman}, {Faber}, {Fisher}, {Fowler},
  {Gallardo}, {Gordon}, {Gralla}, {Hajian}, {Halpern}, {Hasselfield},
  {Hern{\'a}ndez-Monteagudo}, {Hill}, {Hilton}, {Hilton}, {Hincks}, {Holtz},
  {Huffenberger}, {Hughes}, {Hughes}, {Infante}, {Irwin}, {Jacobson},
  {Johnstone}, {Baptiste Juin}, {Kaul}, {Klein}, {Kosowsky}, {Lau}, {Limon},
  {Lin}, {Louis}, {Lupton}, {Marriage}, {Marsden}, {Martocci}, {Mauskopf},
  {McLaren}, {Menanteau}, {Moodley}, {Moseley}, {Netterfield}, {Niemack},
  {Page}, {Page}, {Parker}, {Partridge}, {Plimpton}, {Quintana}, {Reese},
  {Reid}, {Rojas}, {Sehgal}, {Sherwin}, {Schmitt}, {Spergel}, {Staggs},
  {Stryzak}, {Swetz}, {Switzer}, {Thornton}, {Trac}, {Tucker}, {Uehara},
  {Visnjic}, {Warne}, {Wilson}, {Wollack}, {Zhao}, \& {Zunckel}}]{ACT2013}
{Sievers} J.~L. {et~al.}, 2013, Journal of Cosmology and Astroparticle Physics,
  10, 60

\bibitem[{{Slosar} {et~al}\mbox{.}(2008){Slosar}, {Hirata}, {Seljak}, {Ho}, \&
  {Padmanabhan}}]{SlosarHirata2008}
{Slosar} A., {Hirata} C., {Seljak} U., {Ho} S., {Padmanabhan} N., 2008, Journal
  of Cosmology and Astroparticle Physics, 8, 31

\bibitem[{{Slosar}, {Seljak} \& {Makarov}(2004){Slosar}, {Seljak}, \&
  {Makarov}}]{SlosarSeljak2004modeproj}
{Slosar} A., {Seljak} U., {Makarov} A., 2004, Phys.\ Rev.\ D., 69, 123003

\bibitem[{{Strand}, {Brunner} \& {Myers}(2008){Strand}, {Brunner}, \&
  {Myers}}]{Strandbrunner2008}
{Strand} N.~E., {Brunner} R.~J., {Myers} A.~D., 2008, Astrophys.\ J., 688, 180

\bibitem[{{Swanson} {et~al}\mbox{.}(2008){Swanson}, {Tegmark}, {Hamilton}, \&
  {Hill}}]{Swanson2008mangle}
{Swanson} M.~E.~C., {Tegmark} M., {Hamilton} A.~J.~S., {Hill} J.~C., 2008,
  Mon.\ Not.\ Roy.\ Astron.\ Soc., 387, 1391

\bibitem[{{Tegmark}(1997)}]{Teg97}
{Tegmark} M., 1997, Phys.\ Rev.\ D., 55, 5895

\bibitem[{{Tegmark} {et~al}\mbox{.}(2002){Tegmark}, {Dodelson}, {Eisenstein},
  {Narayanan}, {Scoccimarro}, {Scranton}, {Strauss}, {Connolly}, {Frieman},
  {Gunn}, {Hui}, {Jain}, {Johnston}, {Kent}, {Loveday}, {Nichol}, {O'Connell},
  {Sheth}, {Stebbins}, {Szalay}, {Szapudi}, {Vogeley}, {Zehavi}, {Annis},
  {Bahcall}, {Brinkmann}, {Csabai}, {Doi}, {Fukugita}, {Hennessy},
  {Ivez{\'{\i}}c}, {Knapp}, {Lamb}, {Lee}, {Lupton}, {McKay}, {Kunszt}, {Munn},
  {Peoples}, {Pier}, {Richmond}, {Rockosi}, {Schlegel}, {Stoughton}, {Tucker},
  {Yanny}, \& {York}}]{Tegmark2002earlysdss}
{Tegmark} M. {et~al.}, 2002, Astrophys.\ J., 571, 191

\bibitem[{Tegmark {et~al}\mbox{.}(1998)Tegmark, Hamilton, Strauss, Vogeley, \&
  Szalay}]{Tegmark:1997yq}
Tegmark M., Hamilton A.~J., Strauss M.~A., Vogeley M.~S., Szalay A.~S., 1998,
  Astrophys.J., 499, 555

\bibitem[{{Tegmark} {et~al}\mbox{.}(1998){Tegmark}, {Hamilton}, {Strauss},
  {Vogeley}, \& {Szalay}}]{THS1998future}
{Tegmark} M., {Hamilton} A.~J.~S., {Strauss} M.~A., {Vogeley} M.~S., {Szalay}
  A.~S., 1998, Astrophys.\ J., 499, 555

\bibitem[{{Xia} {et~al}\mbox{.}(2011){Xia}, {Baccigalupi}, {Matarrese},
  {Verde}, \& {Viel}}]{Xia2011sdssqsocell}
{Xia} J.-Q., {Baccigalupi} C., {Matarrese} S., {Verde} L., {Viel} M., 2011,
  Journal of Cosmology and Astroparticle Physics, 8, 33

\bibitem[{{Xia} {et~al}\mbox{.}(2009){Xia}, {Viel}, {Baccigalupi}, \&
  {Matarrese}}]{Xia2009highzisw}
{Xia} J.-Q., {Viel} M., {Baccigalupi} C., {Matarrese} S., 2009, Journal of
  Cosmology and Astroparticle Physics, 9, 3

\end{thebibliography}
}
\normalsize

\appendix

\end{document}